 \documentclass[twocolumn,trackchanges]{aastex631}

\usepackage{tikz}
\usetikzlibrary{arrows}
\usepackage{amsmath}

\shorttitle{Spiral shocks}
\shortauthors{Aktar et al.}
\graphicspath{{./}{figures/}}

\begin{document}

\title{Point-wise Self-similar Solution for Spiral Shocks in Accretion Disk with Mass Outflow in Binary}

\correspondingauthor{Ramiz Aktar \& Li Xue}
\email{ramizaktar@gmail.com, lixue@xmu.edu.cn}

\author[0000-0002-3672-6271]{Ramiz Aktar}
\affiliation{Department of Astronomy, Xiamen University, Xiamen, Fujian 361005, People’s Republic of China}

\author[0000-0002-0279-417X]{Li Xue}
\affiliation{Department of Astronomy, Xiamen University, Xiamen, Fujian 361005, People’s Republic of China}


\author[0000-0001-8678-6291]{ Tong Liu}
\affiliation{Department of Astronomy, Xiamen University, Xiamen, Fujian 361005, People’s Republic of China}


\begin{abstract}

We examine the properties of spiral shocks from a steady, adiabatic, non-axisymmetric accretion disk around a compact star in binary. We first time incorporate all the possible influences from binary through adopting the Roche potential and Coriolis forces in the basic conservation equations.  In this paper, we assume the spiral shocks to be point-wise self-similar, and the flow is in vertical hydrostatic equilibrium to simplify the study. We also investigate the mass outflow due to the shock compression and apply it to the accreting white dwarf in binary. We find that our model will be beneficial to overcome the {\it ad hoc} assumption of optically thick wind generally used in the studies of the progenitor of supernovae Ia.

\end{abstract}


\section{Introduction} \label{sec:intro}

The accretion-ejection process is one of the most powerful phenomena in the universe, which is ubiquitously observed around protostars, white dwarfs, neutron stars, and black holes. As the fundamental issue, the transfer of angular momentum and mass is always concerned by many studies in past decades \citep{Pringle-81}. In this regard, \cite{Shakura-Sunyaev73} first proposed the ``$\alpha$-disk" model. This model assumes that the viscosity originates from turbulence, which can be excited through magneto-rotational instability of small-scale tangle magnetic field \citep{Balbus-Hawley91}. \citet{Blandford-Znajek77} proposed the purely electromagnetic energy extraction mechanism from the black hole to boost the jet. On the other hand, \citet{Blandford-Payne82} proposed the wind originating mechanism through the large-scale open magnetic field emerging from the disk. These pioneering works are cited in a lot of follow-up studies and constitute prevailing accretion-ejection theory.

As a useful supplement to prevailing theory, the shocks in accretion flow, including the axisymmetric radially standing shocks and non-axisymmetric spiral shocks, have been rigorously investigated in literature over past decades. Series of theoretical works proposed and gradually improved the radially standing shock model \citep{Fukue-87, Chakrabarti89, Lu-etal99, Becker-Kazanas01, Fukumura-Tsuruta04, Chakrabarti-Das04, Sarkar-Das16, Sarkar-etal18, Dihingia-etal18, Dihingia-etal19a, Dihingia-etal19b}. This model has been used to explain the observational spectral states and quasi-periodic oscillations of compact objects \citep{Chakrabarti-Titarchuk95, Molteni-etal96a}. The mass ejection (jets or outflows from disks) due to the shock compression has also been extensively investigated with this model in recent years \citep{Chattopadhyay-Das07, Das-Chattopadhyay08, Kumar-Chattopadhyay13, Aktar-etal15, Aktar-etal17, Aktar-etal19}.

On the other hand, the spiral shock model was first proposed as an effective angular momentum transfer mechanism by \cite{Michel-84} and gradually improved by a series of subsequent works \citep{Sawada-etal86a, Sawada-etal86b, Matsuda-etal87, Spruit-87, Spruit-etal87, Chakrabarti-90b, Livio-Spruit-91, Lanzafame-etal00, Lanzafame-etal01, Molteni-etal01}. The spiral shock is easily produced due to the disk instabilities or tidal perturbations in the non-axisymmetric flow, even in the inviscid flow, which has been confirmed by various simulation studies \citep{Sawada-etal86a, Sawada-etal86b, Savonije-etal94, Rafikov-02, Bisikalo-etal04, Kurbatov-etal14,  Arzamasskiy-Rafikov18}. Simultaneously, observations also confirm the presence of spiral structure and spiral shock in the accretion disk \citep{Steeghs-etal97, Neustroev-Borisov98, Pala-etal19, Baptista-etal20, Lee-etal20}. 

In this paper, we perform the analytical study of non-axisymmetric and stationary solutions with spiral shocks for the accretion disk in an interacting binary system composed of a compact star and a normal star. Our analytical study is carried out under the assumption of self-similarity, which is first proposed by \cite{Spruit-87} and also adopted by subsequent works \citep[]{Vishniac-Patrick89, Chakrabarti-90b, Larson-90, Hennebelle-etal16}. All of these previous works are only valid for the accretion of a single star because of their implements of the Newtonian gravitational potential. However, for the accretion of compact stars in binary, the complicated and non-axisymmetric Roche potential should be considered in the co-rotating frame (see Figure \ref{Figure_1}). This makes the territory of self-similarity shrink to point-wise regions, and the solution becomes point-wise valid. Meanwhile, we also consider the mass outflow due to the compression of spiral shock according to the similar analyses in \cite{Aktar-etal15, Aktar-etal17, Aktar-etal19}. Therefore, our model is an analytical accretion-ejection model for the interactive binary, which includes the effects characterized by the parameters of binary and might be used in further studies of binary evolution \citep{Wang-18}.

We organize the paper as follows. In section 2, we present the description of the model. In section 3, we discuss the solution methodology and present the results in detail. In section 4, we describe the astrophysical application of our model. Finally, we draw the concluding remarks in section 5.

\section{Model Description}

We consider a steady, adiabatic, non-axisymmetric accretion flow around a compact star in a binary system. We adopt the spiral shock model proposed by \citet{Chakrabarti-90b}. Here, the radial and the azimuthal components of momentum are exactly solved. However, the momentum equation in the vertical direction, i.e., in the off-equatorial plane in the disk, is neglected. Also, the accretion flow is assumed to be in vertical hydrostatic equilibrium. Therefore, it is considered to be a model of 2.5-dimensional as disk height is vertically averaged. We also assume the shocks to be self-similar, and the shock conditions are vertically averaged \citep{Matsumoto-etal84}.

\subsection{Governing Equations}

In this paper, we consider the governing equations with cylindrical coordinates on the equatorial plane in the corotating frame of a binary system. The equations are\\
(i) The radial momentum conservation equation:
\begin{equation}\label{radial_eq}
v_r\frac{\partial v_r}{\partial r} + \frac{v_{\phi}}{r} \frac{\partial v_r}{\partial \phi} +\frac{1}{\rho} \frac{\partial P}{\partial r} - \frac{v_{\phi}^2}{r}- 2\omega v_\phi + \frac{\partial \Phi_R}{\partial r} =0,
\end{equation}
(ii) The azimuthal momentum equation:
\begin{equation}\label{azimuthal_eq}
v_r \frac{\partial v_\phi}{\partial r} + \frac{v_\phi}{r} \frac{\partial v_\phi}{\partial \phi} + \frac{v_\phi v_r}{r} + \frac{1}{r \rho} \frac{\partial P}{\partial \phi} + 2 \omega v_r +\frac{\partial \Phi_R}{r \partial \phi}=0,
\end{equation}
(iii) The continuity equation:
\begin{equation} \label{continuity_eq}
\frac{\partial }{\partial r}(hv_r \rho r) + \frac{\partial }{\partial \phi}(h \rho v_{\phi}) =0,
\end{equation}
and finally\\
(iii) The vertical pressure balance equation:
\begin{equation}\label{vertical_eq}
\frac{1}{\rho} \frac{\partial P}{\partial z} =\left( \frac{\partial \Psi_G}{\partial z}\right)_{z<<r}
\end{equation}
where $r$, $\phi$, $v_r$, $v_\phi$, $P$, $\rho$, and $2h$ are the radial coordinate, azimuthal coordinate, radial component of velocity, azimuthal component of velocity, gas pressure, density of the flow, and local vertical thickness, respectively. The $\Phi_R$ in equation (\ref{radial_eq} and \ref{azimuthal_eq}) is the Roche potential that is an equivalent potential combining the gravity and centrifugal force. The fifth term in equations (\ref{radial_eq}) and (\ref{azimuthal_eq}) arises due to the contribution from Coriolis forces in cylindrical coordinate as $2 \vec{\omega} \times \vec{v}= - 2 \omega v_\phi \vec{e}_r + 2\omega v_r \vec{e}_\phi$, where $\omega$ is the angular velocity of the binary system, $\vec{e}_r$ is the radial unit vector, and $\vec{e}_\phi$ is the azimuthal unit vector respectively. In equation (\ref{vertical_eq}), $\Psi_G$ represents the three-dimensional gravitational potential of the binary system (Since both centrifugal force and Coriolis force have no vertical components, their effects don't need to be considered here). We also use the adiabatic equation of state $P=K\rho^{\gamma}$, where $K$ is the measure of the entropy of the flow. $\gamma =1+ \frac{1}{n}$ is the adiabatic index, and $n$ represents polytropic index of the flow.


\begin{figure}
	\begin{center}
		\includegraphics[width=0.5\textwidth]{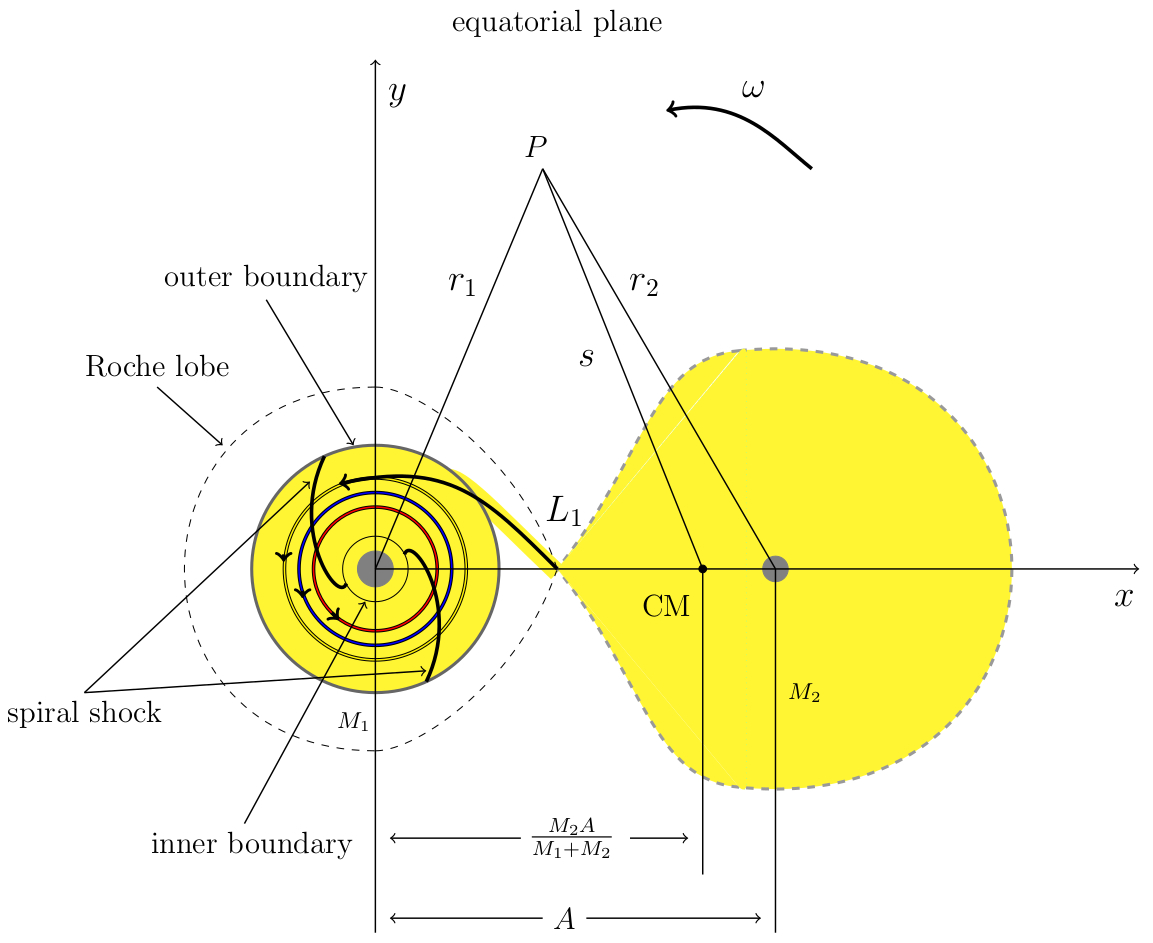} 
	\end{center}
	\caption{Schematic diagram of Roche lobe overflow (RLOF) containing spiral shocks in a binary system.}
	\label{Figure_1}
\end{figure}

\subsection{Roche Potential}

The well-known Roche potential is indispensable for the consideration of hydrodynamics in binary system. In cylindrical coordinate, the three-dimensional Roche potential can be written as
\begin{equation}\label{Roche_3d}
\Psi(r,\phi,z) =  -\frac{GM_1}{r_1} - \frac{GM_2}{r_2} - \frac{1}{2} \omega^2 s^2,
\end{equation}
or
\begin{equation}
\Psi(r,\phi,z) = \Psi_G + \Psi_{CN}.
\end{equation}

Where $G$ is the gravitational constant, $M_1$ and $M_2$ are the mass of primary and secondary stars, $r_1$ and $r_2$ are the distances from centers of two stars, $\omega$ is the rotating angular velocity of binary system, $s$ is the distance from the rotating axis of system, and $\Psi_G$ and $\Psi_{CN}$ are the gravitational and centrifugal potentials respectively. In the co-rotating frame shown in Figure \ref{Figure_1}, these quantities can be calculated by the following formulae
\begin{eqnarray*}
	r_1 &=& \sqrt{r^2+z^2},\\
	r_2 &=& \sqrt{r^2-2r\cos\phi+1+z^2},\\
	\omega &=& \sqrt{1+q} = \omega_0~~{\rm(say)},\\
	s &=& \sqrt{r^2-2 \frac{q}{1+q}r \cos\phi+\left(\frac{q}{1+q}\right)^2},\\
	\Psi_G &=& -\frac{1}{r_1}-\frac{q}{r_2},\\
	\Psi_{CN} &=& -\frac{1}{2}\omega^2 s^2,
\end{eqnarray*} 
where the special unit system of $GM_1 = A = 1$ is adopted here, and $A$ is the separation between two stars, $q (=M_2/M_1)$ is the binary mass ratio. Finally, it is to be noted that the equatorial Roche potential $\Phi_R$ appearing in equations (\ref{radial_eq}) and (\ref{azimuthal_eq}) is easily defined by setting $z=0$ in the formula of $\Psi$ (see equation \ref{Roche_3d}),
and the right hand side of equation (\ref{vertical_eq}) can be easily derived from the formula of $\Psi_G$. 
%

\subsection{Self-similar conditions and flow equations in spiral coordinates}

The governing equations (\ref{radial_eq})-(\ref{vertical_eq}) constitute a system of nonlinear partial differential equations, whose consistent solution can only be obtained by numerical simulation. However, the cost of numerical simulation is huge, and its results are often influenced by many factors (such as computational accuracy, grid resolution, spurious effects in the numerical algorithm, etc.) to become complicated and changeable, so it is difficult to look for insight the simple laws from numerical simulation. Therefore, in order to reduce the computational difficulty, one needs to employ an additional approach about the form of solutions. In this context, the self-similar approach has a tremendous advantage in dealing with the mathematical complexity of a given problem. Self-similarity aims to obtain a dimensionless system of equations, whose essence is to scale physical quantities with local characteristics (e.g., the velocity can be scaled with the Keplerian velocity derived from the primary star, which is read as $r^{-1/2}$ in our special unit system). In this regard, \citet{Spruit-87} first introduces a self-similar approach for steady accretion flow in an inertial frame. Later, \citet{Chakrabarti-90b} studies the spiral shocks in detail by assuming self-similar solutions. On the other hand, \citet{Narayan-Yi94} develop self-similarity conditions for advection-dominated accretion flow. In this work, we also employ the same self-similar approach for spiral shocks but in a co-rotating frame that is a non-inertial frame. It is to be mentioned that the flow equations for accretion are intrinsically two-dimensional using Roche potential. Therefore, we transform the cylindrical coordinate at equatorial plane to spiral coordinate as $(r, \phi) \rightarrow (r, \psi(r, \phi))$, where $\psi = \phi+\beta(r)$ is the spiral coordinate. Following \citet{Spruit-87, Chakrabarti-90b} self-similarity conditions, we write the flow variables as
\begin{align*} 
v_r &= r^{-1/2} q_{1}(\psi), \tag{7a}  \label{self_similar_a}\\
v_\phi &= r^{-1/2} q_{2}(\psi), \tag{7b}\\
a &= r^{-1/2} q_{3}^{1/2}(\psi), \tag{7c}\\
\rho &= r^{-3/2} q_{\rho}(\psi),  \tag{7d}\\ 
P &= r^{-5/2} q_{P}(\psi)  \tag{7e},
\end{align*}
and
\begin{align*}
\frac{\partial \beta}{\partial r} = r^{-1} B \tag{7f} \label{self_similar_f},
\end{align*}
where `spirality' $B = \tan \theta$, $\theta$ is the constant winding angle, i.e., the angle between the radial direction and the outward tangent of the $\psi$ = constant curve. Here, $\theta$ is also known as `pitch angle'.  The measure of entropy $K$ remains constant along the flow in between two shocks, but changes at the shock \citep{Chakrabarti89}. Using the definition of sound speed, we calculate the variation of $K$ as
\begin{equation*}
K = r^{3\gamma/2-5/2}K_0 \tag{8}
\end{equation*}
where, $K_0 = \frac{q_P}{q_\rho}$ \citep{Chakrabarti-90a}. The entropy should increase inward for accretion and outward for wind. Therefore we have $\gamma \le 5/3$ or $n > 3/2$ for accretion and, $\gamma \ge 5/3$ or $n < 3/2$ for wind \citep{Chakrabarti-90a, Chakrabarti-90b}. In this paper, we focus only on the accretion solution, so we choose $n > 3/2$ or $\gamma \le 5/3$ throughout the paper.

Now, from equation (\ref{vertical_eq}), we obtain the disk height $(h)$ as
\begin{equation*} \label{height_eq}
h = \frac{r^{-1/2} q_3^{1/2}}{\mathcal{G}} \tag{9}
\end{equation*}
where, the adiabatic sound speed is defined as $P =\rho a^2$ and $a = r^{-1/2} q_3^{1/2}$. Here, $\mathcal{G}(r, \phi) = \left(\frac{1}{r_1^3} + \frac{q}{r_2^3} \right)^{1/2}$. 
Therefore, using equations (\ref{self_similar_a} - \ref{self_similar_f}) and (\ref{height_eq}), we obtain the dimensionless differential equations of $q_1$, $q_2$ and $q_3$ from equations (\ref{radial_eq} - \ref{vertical_eq}), and are given by
\begin{equation*}\label{dimless_radial_eq}
q_w \frac{d q_1}{d \psi} - \frac{n_\rho + 1}{\gamma} q_3 + \frac{ B}{(\gamma -1)} \frac{dq_3}{d\psi} - \frac{q_1^2}{2} - q_2^2 
  - \frac{2 \omega q_2}{\Omega_K} + \alpha_1 =0,  \tag{10}
\end{equation*}
\begin{equation*}\label{dimless_azimuthal_eq}
q_w \frac{d q_2}{d \psi} +\frac{1}{2} q_1 q_2 + \frac{1}{(\gamma -1)} \frac{d q_3}{d \psi}
  + \frac{2 \omega q_1}{\Omega_K} +\alpha_2= 0, \tag{11}
\end{equation*}
, and
\begin{equation*}\label{dimless_vertical_eq}
B \frac{d q_1}{d \psi} + \frac{d q_2}{d \psi} + \frac{(\gamma +1) q_w}{2(\gamma -1) q_3} \frac{d q_3}{d \psi} - \frac{3}{2} q_1  +\alpha_3 = 0, \tag{12}
\end{equation*}
where
\begin{eqnarray*}
q_w &=& q_2 + Bq_1, \\
\alpha_1 &=& r^2 \frac{\partial \Phi_R}{\partial r}, \\
\alpha_2 &=& r \frac{\partial \Phi_R}{\partial \phi}, \\
\alpha_3 &=& -  \frac{r q_1 }{\mathcal{G}} \left(\frac{\partial \mathcal{G}}{\partial r } \right) - \frac{ q_2}{\mathcal{G}} \left(\frac{\partial \mathcal{G}}{\partial \phi} \right),
\end{eqnarray*}
and $\Omega_K ( = r^{-3/2})$ is the Keplerian angular velocity of primary star. For vertical equilibrium model, we choose $n_\rho =3/2$ \citep{Chakrabarti-90b}. 

It is to be noted that equations (\ref{dimless_radial_eq}) - (\ref{dimless_vertical_eq}) are same as those given by \citep{Chakrabarti-90b} except for some extra terms including $\alpha_1$, $\alpha_2$, $\alpha_3$ and $\Omega_K$. The existence of these terms makes the territory of self-similarity shrink to point-wise, because they are functions of $r$ and $\phi$, which leads to the solutions of $q_1$, $q_2$ and $q_3$ changing with different positions. Therefore, the solution of equations (\ref{dimless_radial_eq}) - (\ref{dimless_vertical_eq}) is only valid in the point-wise region, in which those extra terms are all changeless. So we call it \emph{point-wise self-similar solution} in this paper.

Reviewing the classical self-similar solution, it is a common feature that the Newtonian gravitational potential has been adopted in \cite{Spruit-87}, \cite{Chakrabarti-90a} and \cite{Narayan-Yi94}, which makes their dimensionless flow equations independent of position to maintain the self-similarity of their solutions in a wide territory. However, it also makes the interesting physical setting impossible, e.g., the non-inertial effects from the co-rotating frame of binary. In addition, by comparing with simulations, \cite{Narayan-Yi94} points out that the self-similar solution is only valid in the middle radial region of the accretion disk, in which there is less effect from the inner and outer boundaries, although it can be applied to all available radii mathematically. This shows that the self-similar solution is only a local solution under the local simplification but not a global solution. Therefore, we consider the additional physical effects from the companion gravity, centrifugal force, and Coriolis' force in our model to study the influence of binary system on the accretion flow, and we continue to use the self-similar condition to simplify the calculation and obtain the point-wise valid solution. It will make our \emph{point-wise self-similar solution} contain more physical connotations than the classical self-similar solution.

Due to the changeability of equations (\ref{dimless_radial_eq})-(\ref{dimless_vertical_eq}) with different positions, our strategy is to solve the equations at a series of points with different radii on the spiral shock surface (as shown in Figure \ref{Figure_1}), instead of solving only once for the whole shock surface as \cite{Chakrabarti-90b}. The solutions at these points are combined to form a complete solution on the whole shock surface, which includes the radial variation both from the self-similar condition and those additional terms along the spiral shock surface. In fact, it is closer to the physical reality than the classical self-similar solution.


\subsection{Sonic point conditions}

Now, the sonic point conditions can be obtained by eliminating $\frac{dq_1}{d \psi}$ and $\frac{dq_2}{d \psi}$ from equation (\ref{dimless_vertical_eq}) using equation (\ref{dimless_radial_eq}) and (\ref{dimless_azimuthal_eq}), and is given by
$$
\frac{d q_3}{d \psi} = \frac{N}{D}
\eqno(13)
$$
where,
\begin{align*}
N & =    -\frac{(n_\rho +1) B q_3}{\gamma} - \frac{B q_1^2}{2} - Bq_2^2 - \frac{2 \omega  q_2 B}{\Omega_K}  + B \alpha_1 + \frac{q_1 q_2}{2}   \\ & + \frac{2 \omega q_1}{\Omega_K} + \alpha_2 + \frac{3}{2} q_w q_1 - q_w \alpha_3   \tag{14}
\end{align*}
and,
$$
D = -\frac{B^2}{(\gamma -1)} - \frac{1}{(\gamma -1)} + \frac{q_w^2 (\gamma + 1)}{2(\gamma -1)q_3}.
\eqno(15)
$$

During accretion, the denominator $(D)$ at equation (13) becomes zero at some surfaces, known as sonic surface $\psi=\psi_c$. The numerator $(N)$ is also simultaneously zero at sonic surfaces. The vanishing condition of denominator $D=0$ implies the sound speed at the sonic surface as
$$
q_{3c} = \frac{q_w^2}{(B^2+ 1)} \frac{(\gamma +1)}{2}.
\eqno(16)
$$
In the presence of shock, the velocity component perpendicular to the shock is
$$
q_{\bot} = \frac{q_2 + B q_1}{ (B^2 + 1)^{1/2}},
\eqno(17)
$$
and velocity component parallel to the shock is given by
$$
q_{\parallel} = \frac{q_1 - B q_2}{ (B^2 + 1)^{1/2}}.
\eqno(18)
$$
Similarly, the vanishing condition of numerator $N=0$ provides the radial velocity $(q_{1c})$ at the sonic surface and is given by
$$
q_{1c} = \frac{- \mathcal{B} \pm \sqrt{\mathcal{B}^2 - 4 \mathcal{A} \mathcal{C}}}{2 \mathcal{A}}
\eqno(19)
$$
where,
\begin{align*}
\mathcal{A} & =  - \frac{B^3 (n_\rho + 1)}{(B^2 + 1)}  \frac{(\gamma +1)}{2 \gamma} + B +  \frac{r B}{ \mathcal{G}} \left(\frac{\partial \mathcal{G}}{\partial r}\right)
\end{align*}
\begin{align*}
\mathcal{B} & = - \frac{ B^2 q_2 (n_\rho +1) }{(B^2 + 1)}  \frac{(\gamma +1)}{\gamma} + 2 q_2 + \frac{2 \omega}{\Omega_K} +\frac{r q_2}{ \mathcal{G}} \left(\frac{\partial \mathcal{G}}{\partial r} \right) 
\\ & + \frac{B q_2}{ \mathcal{G}} \left(\frac{\partial \mathcal{G}}{\partial \phi} \right)
\end{align*}
\begin{align*}
\mathcal{C} & = - \frac{B q_2^2  (n_\rho +1)}{(B^2 + 1)} \frac{(\gamma +1)}{2 \gamma} - B q_2^2
- \frac{2 \omega  q_2 B}{\Omega_K} + B \alpha_1 + \alpha_2 
\\ & +  \frac{q_2^2}{ \mathcal{G}} \left(\frac{\partial \mathcal{G}}{\partial \phi} \right)    
\end{align*}
where, the subscript represents the quantities evaluated at the sonic surface. Therefore, to evaluate the derivative $\frac{d q_3}{d \psi}|_c$ at the sonic surfaces, we apply `l'Hospital rule in equation (13) \citep{Chakrabarti89}. The detailed expression of $\frac{d q_3}{d \psi}|_c$ is given in the appendix A.

\begin{figure*}
	\begin{center}
		\includegraphics[height=8.5cm, width=8.5cm]{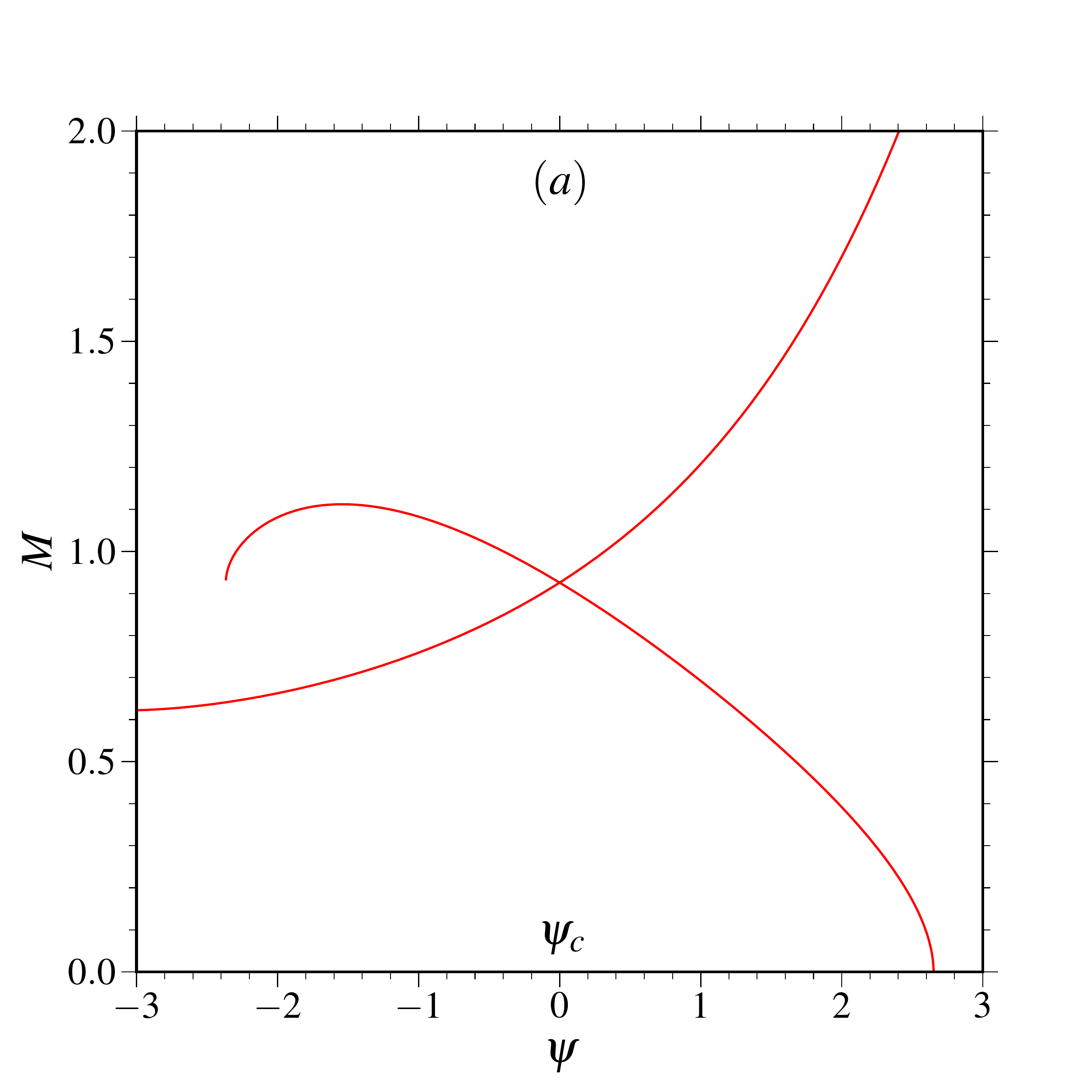} 
		\includegraphics[height=8.5cm, width=8.5cm]{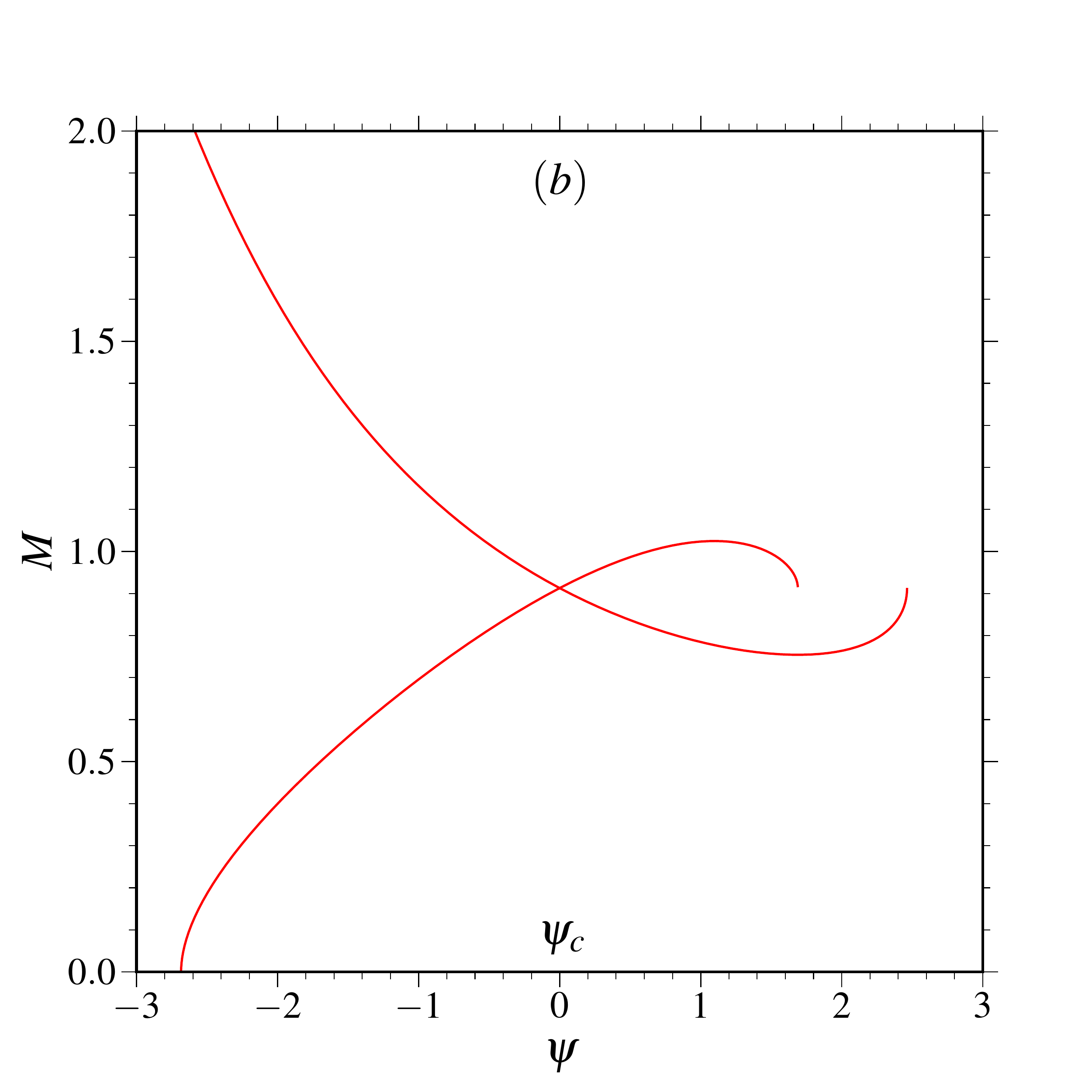} 
	\end{center}
	\caption{Shock free solutions in terms of spiral coordinate. The flow parameters ($\theta, q_{2c}, \gamma$) are (a): ($45^{\circ}$, 0.15, 4/3),  and (b): ($135^{\circ}$, 0.20, 1.4), respectively. The binary parameters $(q, \omega)$ are ($0.1, \omega_0$). Here we fix the radial distance at $r = 0.1$. See the text for details. }
	\label{Figure_2}
\end{figure*}
%
%
\begin{figure*}
	\begin{center}
		\includegraphics[width=0.8\textwidth]{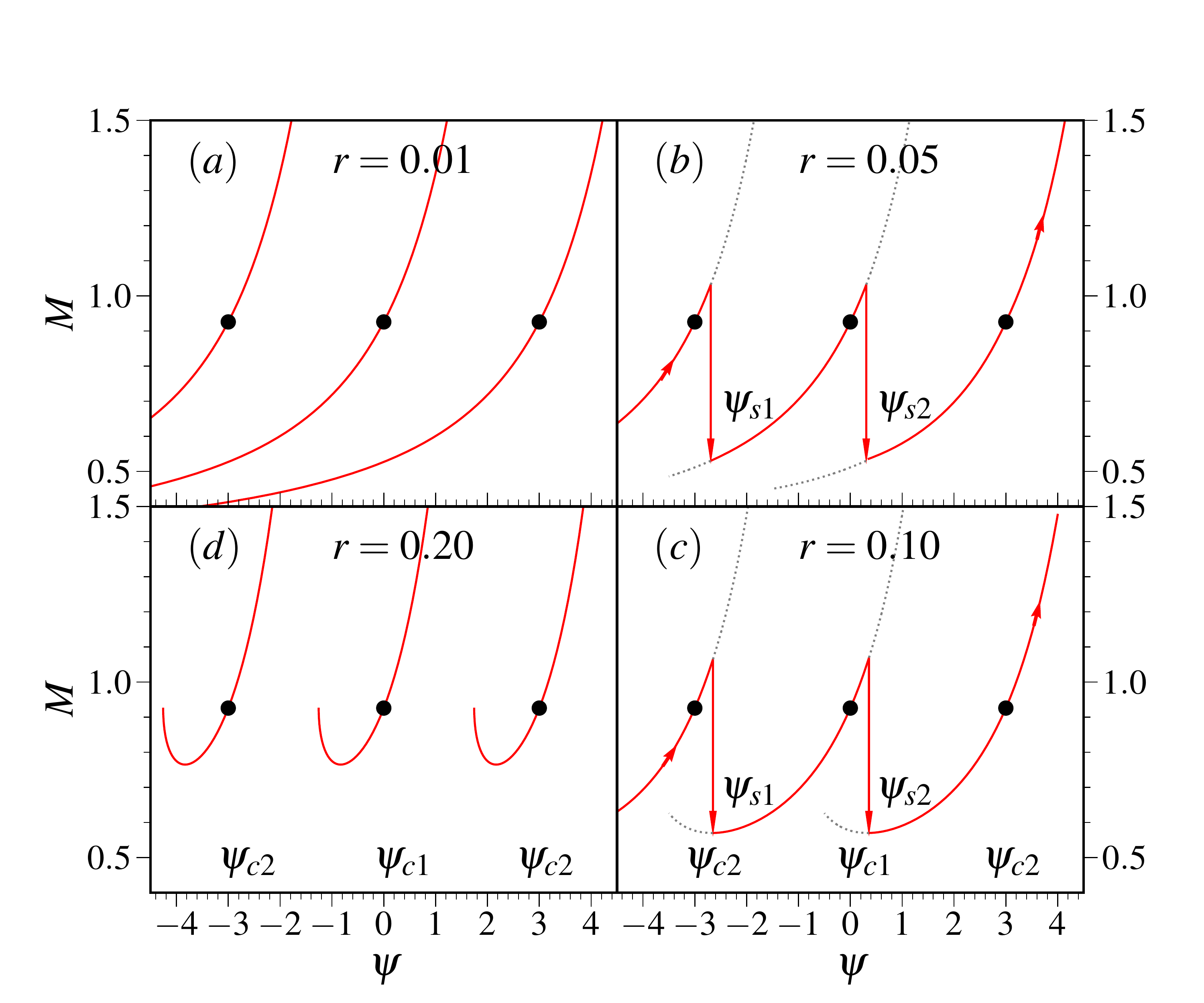} 
	\end{center}
	\caption{Illustration of solution topology for different radial distance. The panel ($a)$ and $(d)$ represent shock free solutions whereas the panel ($b)$ and $(c)$ contain spiral shock waves.  The vertical arrows represent spiral shock transitions in the flow. Here, we fix the flow parameters $(\theta, q_{2c}, \gamma)$ = ($30^\circ, 0.20, 4/3$), and the binary parameters $(q, \omega)$ = (0.1, $\omega_0$). Here, we consider the number of shocks $n_s =2$. See the text for details.}
	\label{Figure_3}
\end{figure*}

\begin{figure}
	\begin{center}
		\includegraphics[height=8.5cm, width=8.5cm]{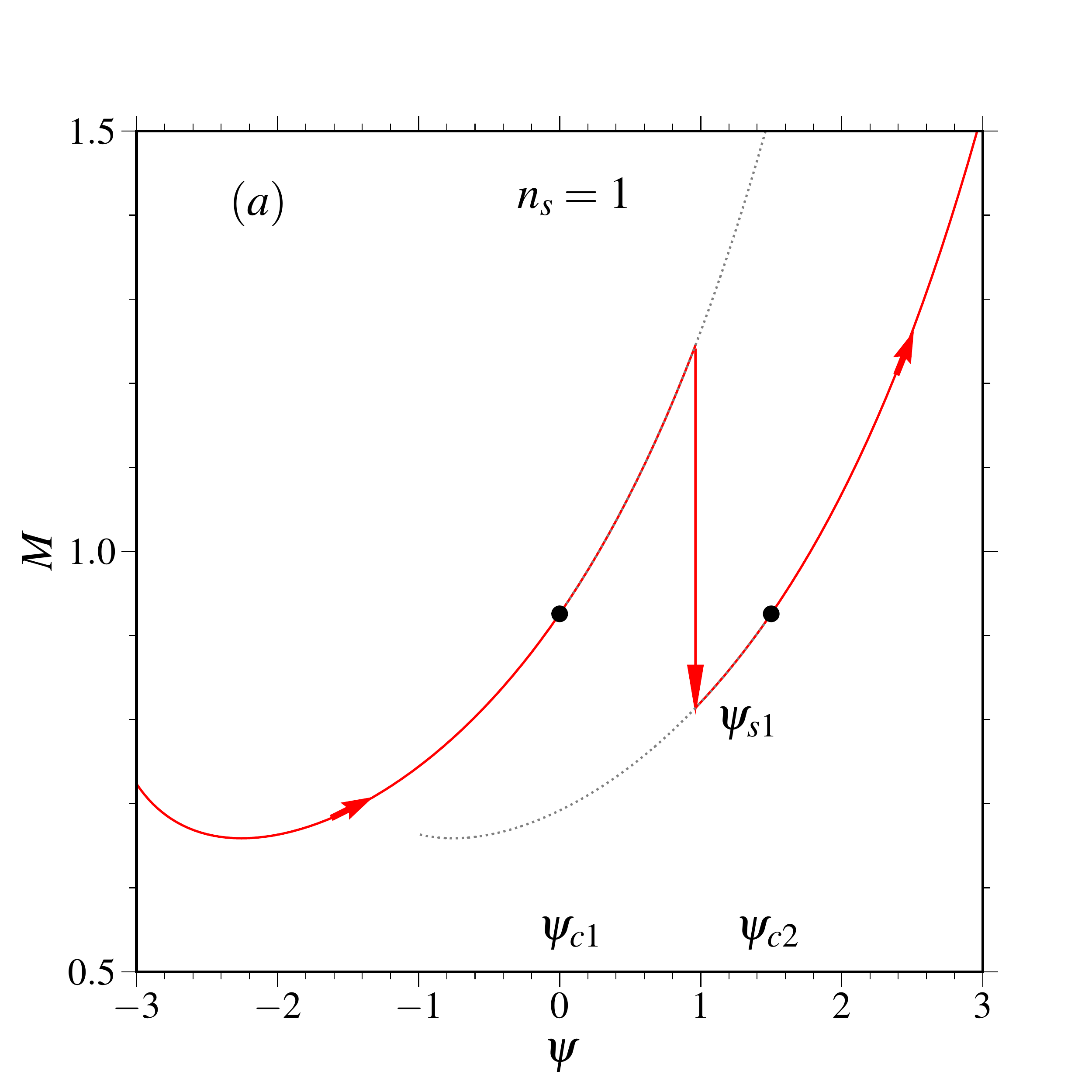} 
		\includegraphics[height=8.5cm, width=8.5cm]{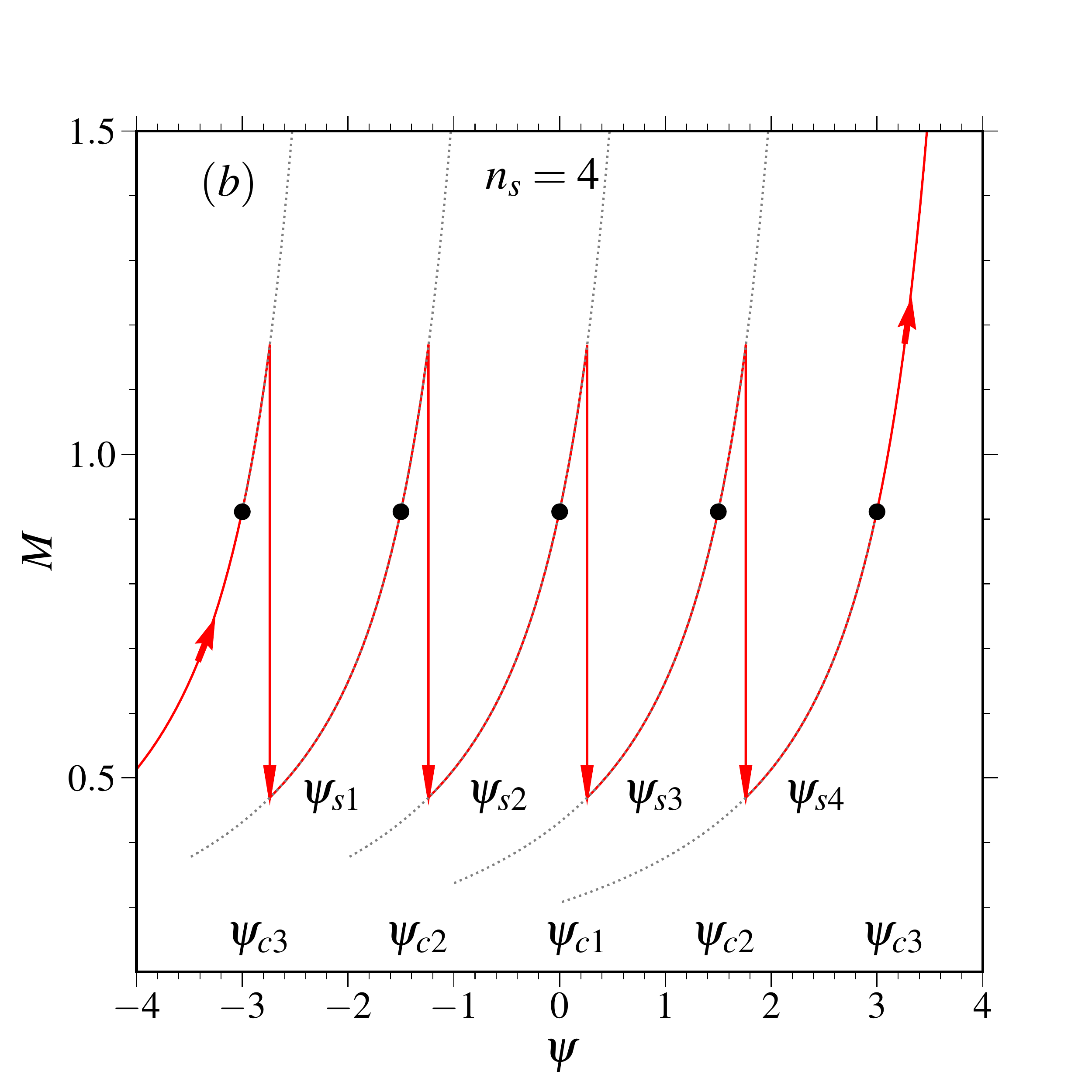} 
	\end{center}
	\caption{Representation of spiral shocks transitions for the number of shocks $n_s =1$ in panel (a) and $n_s =4$ in panel (b), respectively. The vertical arrows represent spiral shock transitions in the flow. Here, the  flow parameters $(\theta,q_{2c}, \gamma, r)$ are for panel (a): ($45^\circ$, 0.12, 4/3, 0.15) and panel (b):($15^\circ$, 0.15, 1.4, 0.1), respectively. We also fix the binary parameters as $(q, \omega)$ = (0.1, $\omega_0$) for both the cases. See the text for details.}
	\label{Figure_4}
\end{figure}

%
\begin{figure*}
	\begin{center}
		\includegraphics[width=0.8\textwidth]{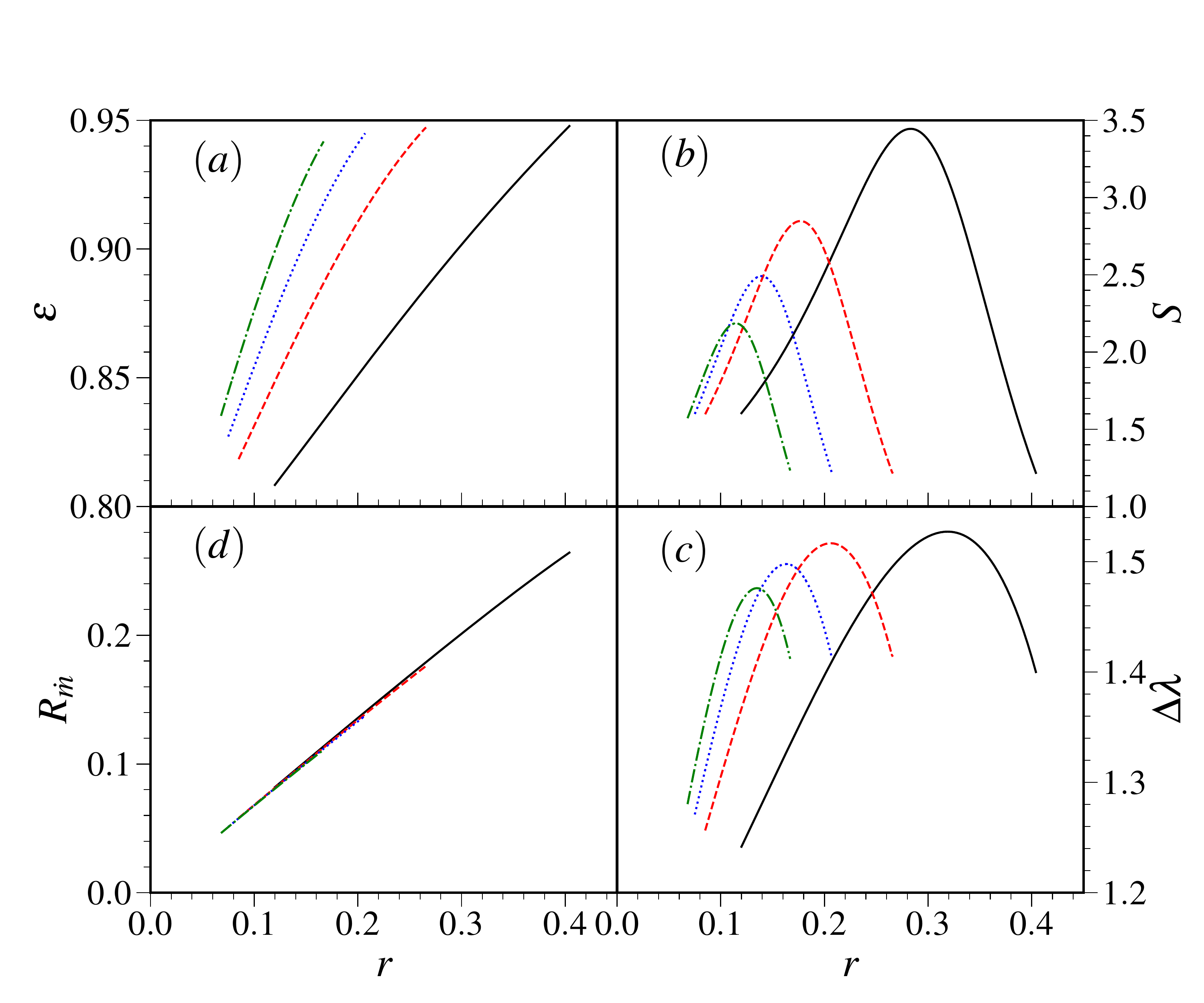} 
	\end{center}
	\caption{Variation of $(a)$: shock locations $(\epsilon)$, $(b)$: shock strength $(\mathcal{S})$, $(c)$: amount of angular momentum dissipation $(\Delta \lambda)$ across shock, and $(d)$: mass outflow rate $(R_{\dot{m}})$ in terms of radial distance $(r)$ for various angular velocity ($\omega$). The solid (black), dashed (red), dotted(blue), and  dashed-dotted (green) curves are for $\omega$ = 0.5, $\omega_0$, 1.5, and 2.0, respectively. Here, we fix $\theta=30^{\circ}$, $q_{2c} = 0.10$, $q =0.10$, and $\gamma= 4/3$. See the text for details.}
	\label{Figure_5}
\end{figure*}

\begin{figure*}
	\begin{center}
		\includegraphics[width=0.8\textwidth]{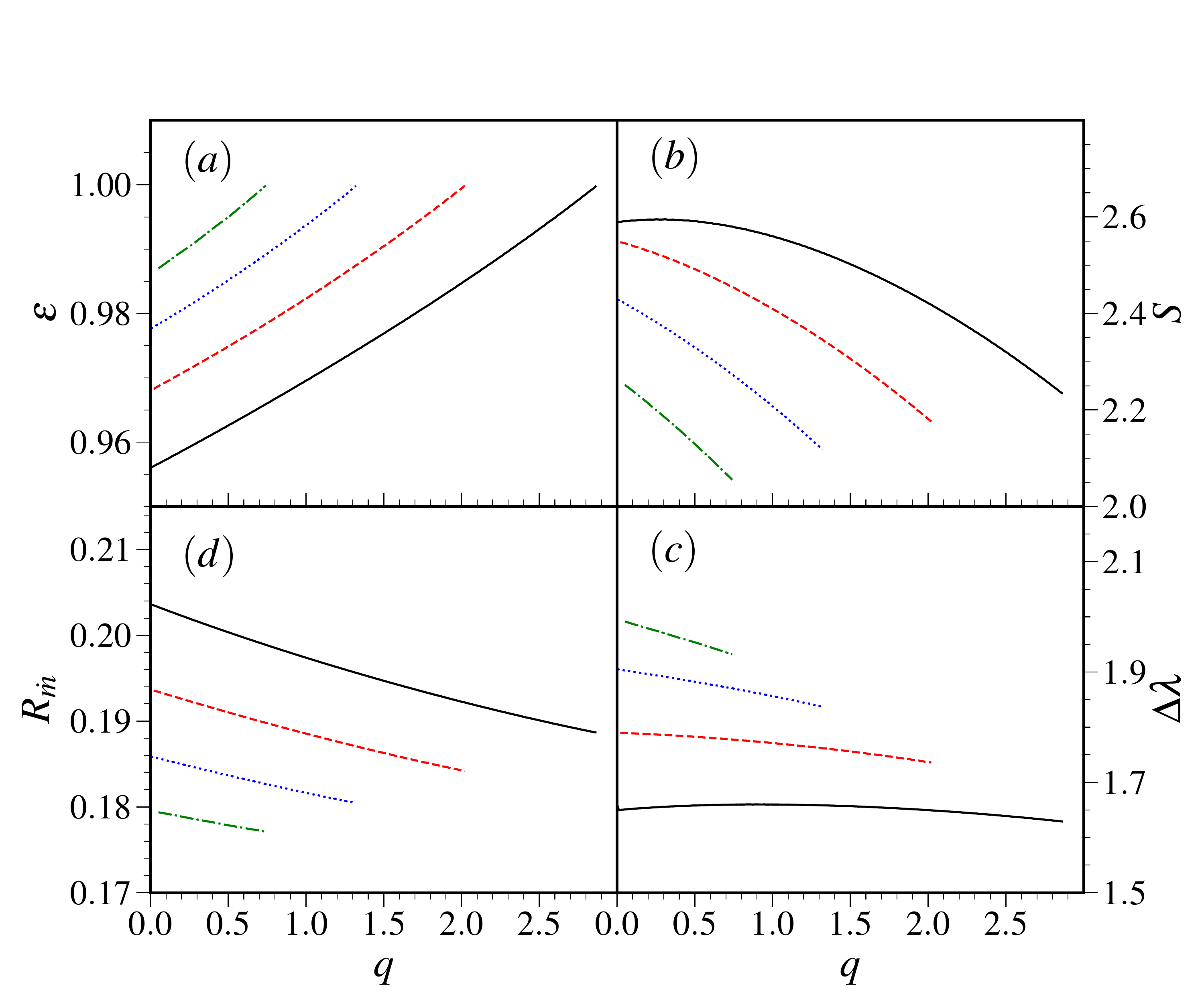} 
	\end{center}
	\caption{Variation of $(a)$: shock locations $(\epsilon)$, $(b)$: shock strength $(\mathcal{S})$, $(c)$: amount of angular momentum dissipation $(\Delta \lambda)$ across shock, and $(d)$: mass outflow rate $(R_{\dot{m}})$ in terms of mass ratio $(q)$ for various angular velocity ($\omega$). The solid (black), dashed (red), dotted(blue), and  dashed-dotted (green) curves are for $\omega$ = 0.25, 0.5, 0.75, and 1.0, respectively. Here, we fix $\theta=35^{\circ}$, $q_{2c} = 0.20$, $r =0.25$, and $\gamma= 4/3$. See the text for details.}
	\label{Figure_6}
\end{figure*}
%

\begin{figure*}
	\begin{center}
		\includegraphics[width=0.49\textwidth]{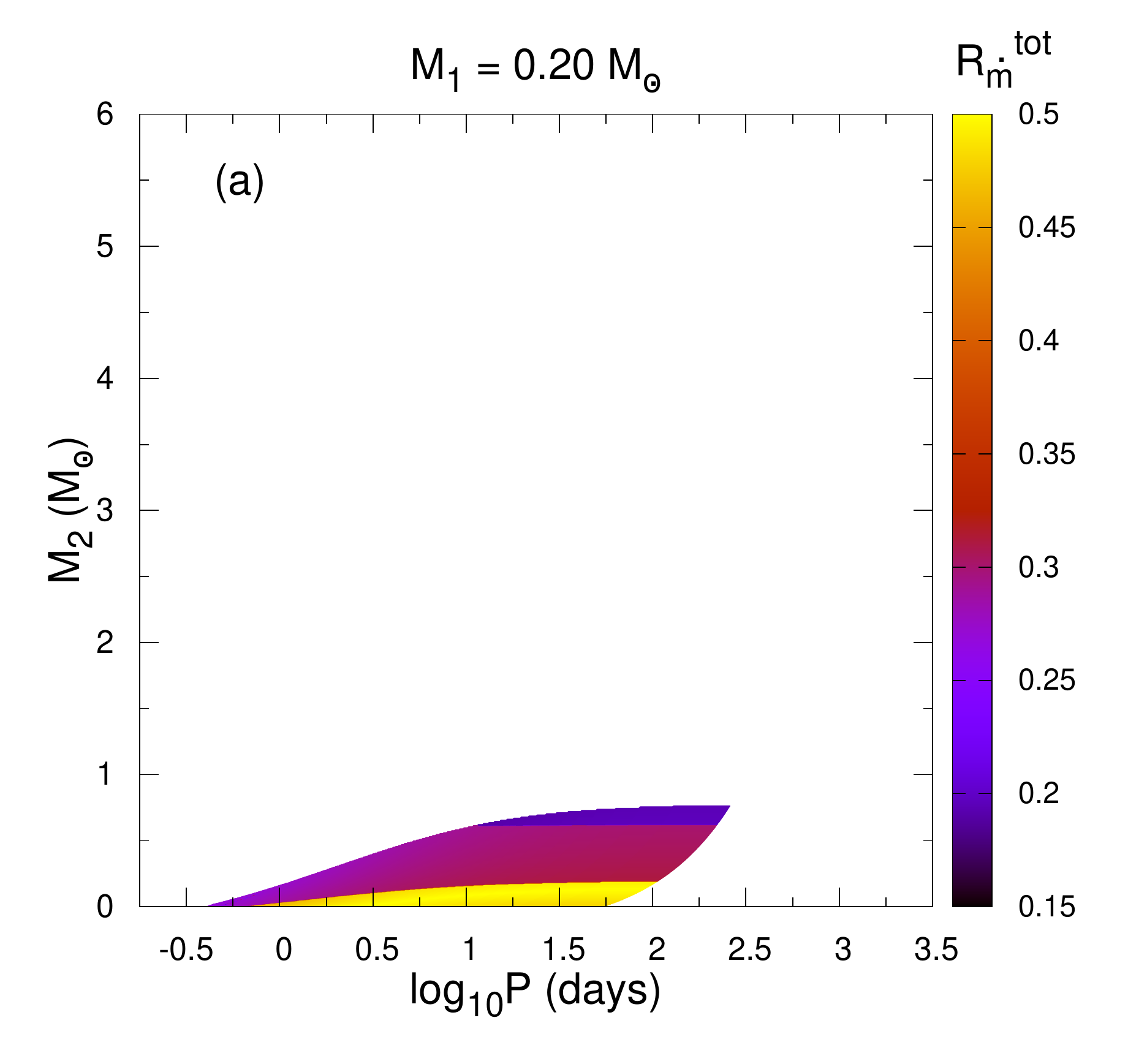}  
		\includegraphics[width=0.49\textwidth]{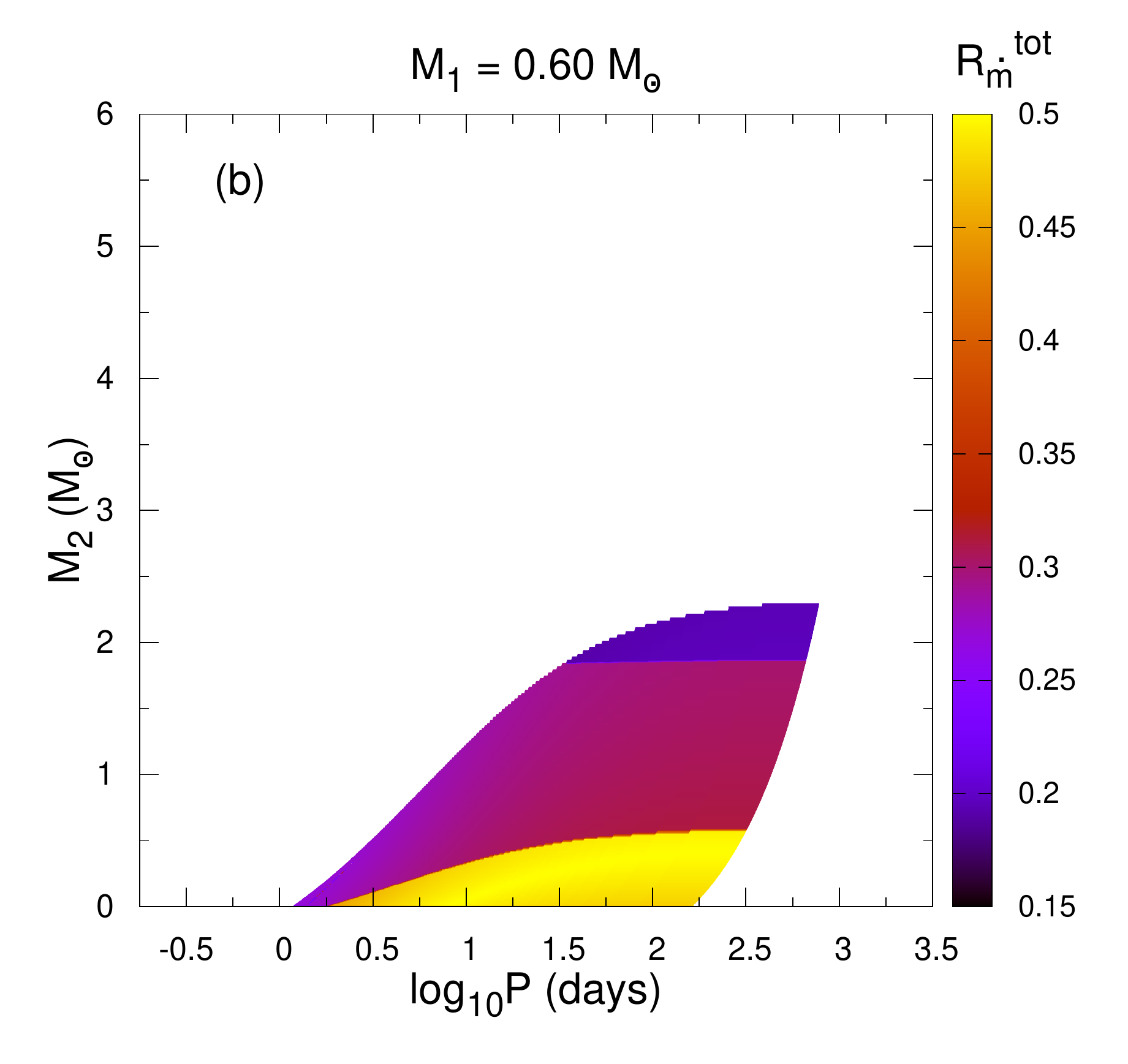}
		\includegraphics[width=0.49\textwidth]{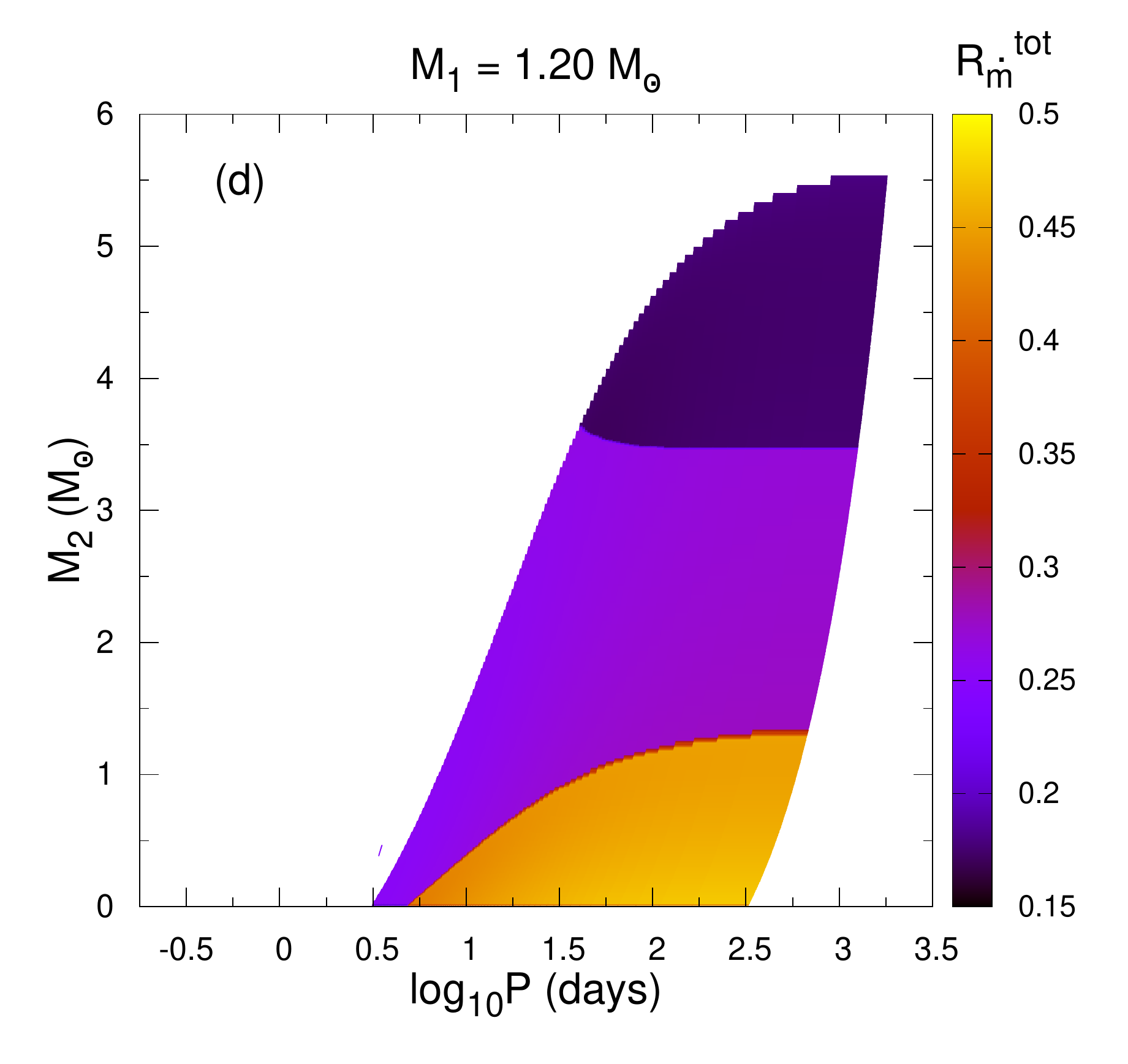}
		\includegraphics[width=0.49\textwidth]{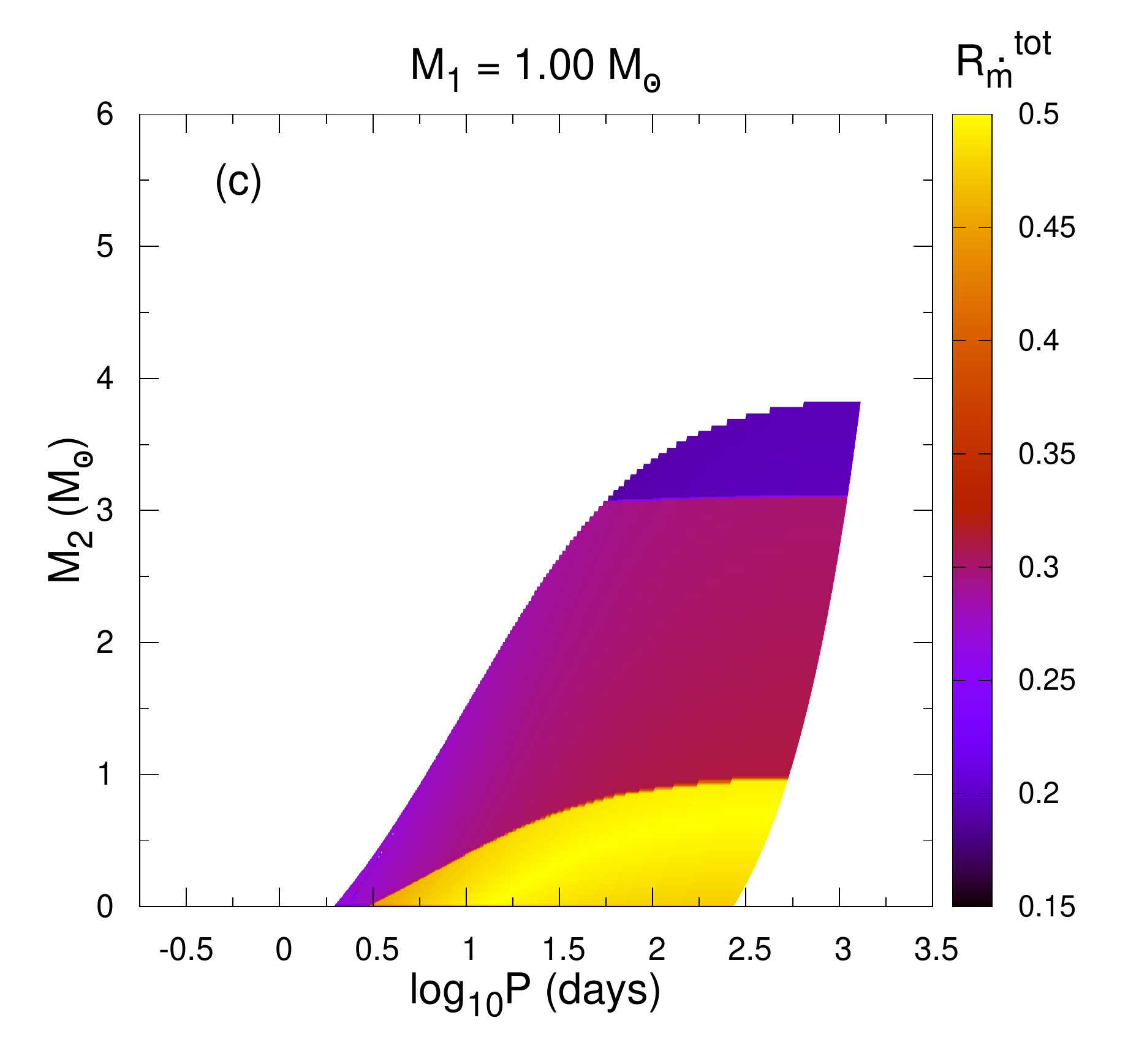}
	\end{center}
	\caption{Two-dimensional surface projection of the three-dimensional plot of orbital period of the binary ($\log_{10}P$) in days - mass of the secondary star $(M_2)$ in terms of total mass outflow rates $(R_{\dot{m}}^{\rm tot})$ by fixing three radial distances i.e. $r=0.15, 0.25$ and 0.35, respectively. In each panel, vertical color coded bar represents the total mass outflow rates. Here, we choose mass of WD as $M_1 =0.2, 0.6, 1.0$ and $1.20 M_{\odot}$. We fix other parameters as $(\theta, q_{2c}, \gamma): (35^{\circ}, 0.20, 4/3)$. See the text for details. }
	\label{Figure_7}
\end{figure*}
%
%

\subsection{In the limiting case: for a single star}
Interestingly, our generalized model for binary system exactly reduce to \citet{Chakrabarti-90b} model in the limiting case for a single star. Considering single star, we choose $M_2 \rightarrow 0$, $\omega \rightarrow 0$ and $G M_1=1$ in the limiting case. Therefore, we obtain from equation (5)
\begin{align*}
& \Phi_R = -\frac{1}{r}, \frac{\partial \Phi_R}{\partial r} = \frac{1}{r^2}, \frac{\partial \Phi_R}{\partial \phi} = 0,
\\& {\rm and}
\\ &  \mathcal{G} = \frac{1}{r^{3/2}}, \frac{\partial \mathcal{G}}{\partial r} = -\frac{3}{2}\frac{1}{r^{5/2}} ,
\frac{\partial  \mathcal{G}}{\partial \phi} = 0. \tag{20}
\end{align*}
Substituting these limiting values in the basic conservation equations (10-12), we retain the same conservation equations mentioned in \citet{Chakrabarti-90b} paper. Therefore, it is clear that the model of \cite{Chakrabarti-90b} is just a limiting case of our generalized model.

\subsection{Analytical expression for shock location}
In this work, we assume that the formation of shocks $(\psi_s)$ are at equidistant in the angular scale $\psi$, and the sonic surfaces $(\psi_c)$ lie in between the shocks are also equidistant \citep{Chakrabarti-90b}. If there is $n_s$ number of shocks present in the flow, the angular separation between two successive shocks and the sonic surfaces is given by $\delta \psi = 2\pi/n_s$. Thus, we can write the shock location formed just prior to a sonic surface $(\psi_{c1})$ as, $\psi_{s1} = \psi_{c1}-\epsilon \delta \psi$ and, location of a shock formed just after the sonic surface as, $\psi_{s2} = \psi_{c1}+(1-\epsilon) \delta \psi$, where $0< \epsilon<1$. Here, $\epsilon$ can be determined by supplying the conserved quantity of the flow \citep{Chakrabarti-90b}. Now, the shock conditions are given by

(1) The energy conservation across the shock:
$$
\frac{q_{3+}}{\gamma -1} + \frac{q_{\bot +}^2}{2} = \frac{q_{3-}}{\gamma -1} + \frac{q_{\bot -}^2}{2}
\eqno(21a)
$$

(2) The momentum conservation across the shock:
$$
W_+ + \Sigma_{+} q_{\bot +}^2 = W_- + \Sigma_{-} q_{\bot -}^2
\eqno(21b)
$$

(3) The conservation of mass flux normal to the shock:
$$
h_+ ~q_{\rho +}~ q_{w +} = h_-~ q_{\rho -}~ q_{w -}
\eqno(21c)
$$

(4) The conservation of velocity component parallel to the flow:
$$
q_{1+} - Bq_{2+} = q_{1-} - Bq_{2-}
\eqno(21d)
$$
where $\pm$ implies post-shock and pre-shock quantities, respectively. $W$ and $\Sigma$ represent vertically integrated gas pressure and density of the flow \citep{Matsumoto-etal84, Chakrabarti89}. Using equations (21a -21c), we obtain the so-called shock invariant quantity $(C_s)$ at the shock and is given by \citep{Chakrabarti89}
$$
C_s = \frac{\left[M_{+} (3\gamma -1) + \frac{2}{M_+}\right]}{\left[2 + (\gamma -1)M_{+}^2\right]} = \frac{\left[M_{-} (3\gamma -1) + \frac{2}{M_-}\right]}{\left[2 + (\gamma -1)M_{-}^2\right]}
\eqno(22)
$$
where $M = \frac{q_\bot}{a}$ is the Mach number of the flow component normal to the shock. To obtain the analytical expression of shock location $\epsilon$, we perform Taylor expansion of the variables $q_1$, $q_2$ and $a$ around the sonic surface $(\psi_c)$ up to second order in $\delta \psi$ following \citet{Chakrabarti-90b}, and is given by
$$
\epsilon = \frac{1}{2} + \frac{1}{\delta \psi} \frac{\left(\frac{d q_{\parallel}}{d \psi}\right)_c}{\left(\frac{d^2 q_{\parallel}}{d \psi^2}\right)_c},
\eqno(23)
$$
where the calculation of the second-order derivatives at the sonic surfaces are shown in the appendix B. If the number of shocks approaches infinity ($n_s \rightarrow \infty$), $\left(\frac{d q_{\parallel}}{d \psi}\right)_c \rightarrow 0$. It immediately implies that $\epsilon \rightarrow \frac{1}{2}$ from equation (23). Here, we define the shock strength as $\mathcal{S} =\frac{M_-}{M_+}$. Also, the amount of specific angular momentum $(\lambda)$ dissipated in presence of the spiral shocks is 
$$
\Delta \lambda = \frac{\lambda_+}{\lambda_-} = \frac{q_{2+}}{q_{2-}}.
\eqno(24)
$$
It allows us to measures the transport of angular momentum due to spiral shock.

\subsection{Mass outflow rate from the disk}

Further, the mass accretion rate in the radial direction can be obtained following \citet{Chakrabarti-90b} for the vertical equilibrium model as
$$
\dot{M}_{\rm in} = \int_0^{2 \pi}{q_1 q_{\rho}q_3^{1/2}} d\psi.
\eqno(24)
$$
On the other hand, the rate of outflowing matter from the disk can be obtained by estimating the accretion rate normal to shock, i.e.,
$$
\dot{M}_{\rm out} = \int_0^{2 \pi}{h q_{\rho} q_{w }} d\psi.
\eqno(25)
$$
Therefore, the mass outflow rates can be obtained as $R_{\dot{m}}= \frac{\dot{M}_{\rm out} }{\dot{M}_{\rm in} }$.

\subsection{Nature of sonic surfaces and solution methodology}

Before going to discuss the solution topology, we first need to determine the nature of sonic surfaces. The discriminant $\mathcal{D} = Q_2^2 - 4Q_1 Q_3$ from equation (A4) in appendix A must be greater than zero $(\mathcal{D}>0)$ for physical sonic surfaces. Based on the theory of critical points, the sonic surfaces are `saddle' type or `X' type if $Q_3<0$; `straight-line if $Q_3 =0$ and, `nodal' type if $Q_3>0$ \citep{Chakrabarti-90a}. The sonic surfaces are `spiral' type or `unphysical' if discriminant $\mathcal{D}<0$. The technical term `spiral' in sonic point analysis implies that the derivative $\frac{dq_3}{d \psi}|_c$ is complex (see equation A4 in appendix A), and it has nothing to do with the spiral flow patterns in the physical space that we are interested in this paper. The nature of the sonic surfaces is solely determined based on the flow parameters, namely pitch angle, azimuthal velocity, adiabatic index $(\theta, q_{2c}, \gamma)$ and binary parameters, namely binary mass ratio and angular velocity $(q, \omega)$, respectively. There are mainly two types of solution; accretion: the matter is spiraling inward, and wind flow: the matter is spiraling outward \citep{Spruit-87, Chakrabarti-90b}.

Here, we briefly describe the solution methodology. The solution method is as follows: We first fix the azimuthal angle ($\phi$) and radial distance $(r)$. After determining ($r, \phi$), the problem is well defined. In this work, we choose $\phi =0$ throughout our calculation; otherwise, it is stated. As a result, the assumption of self-similarity in the radial direction is retained in our model. Now to obtain the solution, we supply the number of shocks $(n_s)$, pitch angle $(\theta)$, rotational velocity at the sonic surface $(q_{2c})$, and adiabatic index ($\gamma$) of the flow. Then, using equations (22) and (23), we self-consistently determine the shock location $(\epsilon)$. It is to be mentioned that the shock locations calculated using analytical expression do not match with the numerically obtained shock location. This is due to the second-order approximation in obtaining $\epsilon$ \citep{Chakrabarti-90b}. It is obvious that the higher-order approximation leads to the actual shock locations. Nevertheless, it is possible to confirm the existence of spiral shocks in the flow from our analytical model even though the shock locations may not be the actual ones.

\section{Results}

In a binary system, the mass is transferred from the donor star through the inner Lagrange point $L_1$ via Roche lobe overflow (RLOF) towards the compact primary star. Then the matter is gravitationally accreted by the primary star. During accretion, matter encounters spiral shocks and transfers angular momentum outwards to reach the primary star surface. A part of the accreting matter is driven out from the disk surface as mass outflow due to the excess thermal gradient force or shock compression at the shock surfaces (see \citet{Aktar-etal15, Aktar-etal17}). The schematic diagram of RLOF in the presence of spiral shock is shown in Figure \ref{Figure_1}. In our model, the solution can be obtained at different radial positions. In order to obtain the solution topology, we supply the flow parameters $\theta$, $q_{2c}$, and $\gamma$ at the sonic surface for a particular radial position, and we fix the binary parameters $(q, \omega)$. Then we integrate numerically using equations (10-12) considering both the slope of equation (A4) in appendix A. It is obvious that not all the flow parameters exhibit spiral shocks. So, we first investigate the shock-free solution for accretion in Figure \ref{Figure_2}a and Figure \ref{Figure_2}b, respectively. To represent the phase space behavior in terms of spiral coordinate, we plot Mach number ($M$) in terms of spiral coordinate ($\psi$) at a particular radial distance. The flow parameters ($\theta, q_{2c}, \gamma$) for Figure \ref{Figure_2}a and Figure \ref{Figure_2}b are $(45^\circ, 0.15, 4/3)$ and $(135^\circ, 0.20, 1.4)$, respectively. The `saddle' type sonic surface is represented by $\psi_c$. We obtain two branches at $\psi_c$ due to the existence of two solutions for $\frac{d q_3}{d \psi}|_c$ at the sonic surfaces (see equation A4). It is observed that the two branches approach towards `spiral' sonic surfaces, and therefore, the branches are `unphysical' in nature. Therefore, the shock conditions are not satisfied for these two cases (Figures \ref{Figure_2}a and \ref{Figure_2}b). As a matter of fact, the shock-free solutions (Figure \ref{Figure_2}a and 2b) fall into the central object after encircling once around the compact primary star to maintain the self-similarity conditions. It is to be noted that in our model, Mach number $(M)$ is drawn for a fixed radial distance $r$ in Figure 2a and Figure \ref{Figure_2}b. Here, we fix the radial distance for Figure \ref{Figure_2}a and Figure \ref{Figure_2}b is at $r = 0.1$. The value of Mach number at the sonic surface is equal to $ M_c = \sqrt{\frac{2}{\gamma +1}}$ (see equation 16 and 17) which is same for axisymmetric vertical equilibrium model \citep{Chakrabarti89}. We also fix the binary parameters as $(q, \omega)$ = $(0.1, \omega_0)$. 

It is obvious that the nature of accretion flow changes at different radial distance during accretion towards the primary star. The major advantage of our {\it point-wise self-similar} solution model is that we can identify the nature of the flow in different radial distance even though we are dealing with spiral coordinate. In this regard, we investigate the nature of accretion flow at different radial distance. Thus, we plot mach number of the accretion flow in terms of spiral coordiates at different radial positions in Figure \ref{Figure_3}. For the purpose of representation, we fix the radial distance at $r = 0.01$, 0.05, 0.1, and 0.2 in panel $(a)$, $(b)$, $(c)$, and $(d)$, respectively. The other flow parameters ($\theta, q_{2c}, \gamma$) = $(30^\circ, 0.20, 4/3)$ and binary parameters $(q, \omega)$ = $(0.1, \omega_0)$ are kept fixed for all the cases. We observe that the flow contains shock free solutions at very near to primary star (panel $(a)$) as well as far away from the star (panel $(d)$). In between, the flow exhibits spiral shock waves where the number of shocks is $n_s =2$ for both panel $(b)$ ($r =0.05$), and $(c)$ ($r =0.1$). The black solid circles represent saddle type sonic surfaces, and arrows indicate flow direction. During accretion, the matter passes through sonic surfaces to become supersonic, and if the shock conditions are satisfied then the flow jumps discontinuously from supersonic branch to subsonic branch to become subsonic. Again the subsonic flow crosses another sonic surface, and the shock transition happens again. Ultimately, the flow hits the primary star surface after spiraling as the matter looses its angular momentum due to shock transition. The vertical arrows represent the spiral shock transitions in the flow. Here, the shock parameters are for panel $(b)$:  $\epsilon = 0.9021$, $M_- = 1.0315$ and $M_+ = 0.5302$, and $\mathcal{S} =1.9454$, and $(c)$:  $\epsilon = 0.8857$, $M_- = 1.0688$ and $M_+ = 0.5700$, and $\mathcal{S} =1.8750$, respectively. The sonic surfaces ($\psi_{c1}, \psi_{c2}$) and shock locations ($\psi{s1}, \psi_{s2}$) are also indicated in the figure.  Similarly, we also show the solution with spiral shocks for number of shocks $n_s =1$ and $n_s =4$, depicted in Figure \ref{Figure_4}a and \ref{Figure_4}b, respectively. Here, we fix the parameters ($\theta, q_{2c}, \gamma, r$) as for Figure \ref{Figure_4}a: $(45^\circ, 0.12, 4/3, 0.15)$, and Figure \ref{Figure_4}b: ($15^\circ$, 0.15, 1.4, 0.1), respectively. The corresponding shock parameters $(\epsilon, M_-, M_+, \mathcal{S})$ are for Figure \ref{Figure_4}a: (0.846816, 1.245271, 0.813942, 1.5299) and Figure \ref{Figure_4}b: (0.8348, 1.1684, 0.4692, 2.4901), respectively. The sonic surfaces ($\psi_{c1}, \psi_{c2}, \psi_{c3}$) and shock locations ($\psi_{s1}, \psi_{s2}, \psi_{s3}, \psi_{s4}$) are shown in Figure \ref{Figure_4}.  Here, the  binary parameters are fixed at $(q, \omega)$ = $(0.1, \omega_0)$ for both Figure \ref{Figure_3} and Figure \ref{Figure_4}. 

Further, we investigate the overall shock properties with the variation of radial distance. In this regard, we plot spiral shock locations in terms of radial distance for fixed flow and binary parameters. We observe that the shock location increases with the increase of radial distance, depicted in Figure \ref{Figure_5}a. The corresponding variation of the strength of shock is shown in Figure \ref{Figure_5}b. It is observed that the shock strength maximizes in the intermediate allowed radial distance. Comparatively weak shocks are formed near to the star surface and far away from the surface. It is evident that strong spiral shock implies more efficient transport of angular momentum outwards. Therefore, we find that the amount of angular momentum dissipated at the shock follows the same trend of shock strength, shown in Figure \ref{Figure_5}c. Finally, we plot the mass outflow rate from the disk in terms of radial distance using equations (24) and (25). It is found that mass outflow rates increase with the increase of radial distance. Also, it is observed that the mass outflow rates lie in the range $\sim 5 \% - 25\%$ for the allowed radial coordinate, depicted in Figure \ref{Figure_5}d. Here, solid (black), dashed (red), dotted (blue), and dashed-dotted (green) are for the angular velocity $\omega$ = 0.5, $\omega_0$, 1.5, and 2.0, respectively. We fix other parameters as $(\theta, q_{2c}, q, \gamma)$ = ($30^\circ$, 0.10, 0.10, 4/3).

Proceed further; we investigate the effect of binary parameters on shock dynamics for fixed flow parameters and at a fixed radial distance. In Figure \ref{Figure_6}a, we represent shock locations in terms of mass ratio $(q)$ for various angular velocity $(\omega)$. We observe that shock location increases with the increase of binary mass ratio for all the cases at a fixed radial distance. Moreover, for fixed $q$, shock location increases with the angular velocity. Figure \ref{Figure_6}b represents the variation of shock strength with mass ratio. It is observed that shock strength decreases with the increase of mass ratio. A similar trend is also observed for the angular momentum dissipated at the shock with the variation of mass ratio, shown in Figure \ref{Figure_6}c. We also observe that shock strength decreases with angular velocity for a fixed mass ratio. Interestingly, we find that transport of angular momentum increases with the increase of angular velocity, i.e., with the increase of centrifugal force. On the other hand, the mass outflow rates decrease with the increase of mass ratio. However, outflow rates increase with angular velocity for a fixed binary mass ratio, depicted in Figure \ref{Figure_6}d. We find that the mass outflow rates lie in the range $\sim 18\%-20\%$. Here, solid (black), dashed (red), dotted (blue), and dashed-dotted (green) are for the angular velocity $\omega$ = 0.25, 0.5, 0.75, and 1.0, respectively. We fix other parameters as $(\theta, q_{2c}, r, \gamma)$ = ($35^\circ$, 0.20, 0.25, 4/3).

\section{Application to white dwarf in a binary system}

Most of the stars form in binary and co-evolve into the interactive state composed of an accreting white dwarf (WD) and a normal star filling Roche lobe. Accreting WD can accumulate material through mild thermonuclear reaction on its surface to grow in mass. If this accretion continues, it is believed that the supernova of Type Ia (SN Ia) will be triggered when the mass of WD finally reaches the Chandrasekhar limit \citep{Hoyle-Fowler60, Nomoto-etal84}. This is the so-called single degenerate (SD) channel for the progenitor of SN Ia. The mass growth of WD in SD channel is very sensitive to the accretion rate. \cite{Wang-18} gives the upper and lower limits of the accretion rate for different massive WD, within which WD can stably grow in mass. Due to the lack of the hydrodynamic study for the WD accretion in SD channel, the inexact artificial method has to be used in the mass exchanging state in evolutionary studies for the progenitor of SN Ia. When the accretion rate exceeds the upper limit, the excess material can only be artificially removed from the WD surface through the assumed optically thick wind, which is not supported by observations \citep{Prieto-etal08, Badenes-etal09, Galbany-etal16, Wang-18}. However, if the accretion-ejection process induced by spiral shock can be considered, the mass finally reaching the WD surface would become reasonable, and the artificial optically thick wind might be reduced or even avoided.

In this section, we apply our model to a binary system in SD channel to exhibit the mass outflow. It is clear that the accretion dynamics are dependent on the binary parameters $\omega$ and $q$ (see Figure \ref{Figure_5} and Figure \ref{Figure_6}). In order to facilitate further comparison, we analyse the parameter space in terms of the orbital period-secondary mass $(P-M_2)$, which is generally used in progenitor studies of SNe Ia \citep{Wang-18} and is also equivalent to $(\omega - q)$ space. For other parameters, we vary the mass of WD as $M_1$ = 0.2, 0.6, 1.0 and 1.20 $M_{\odot}$ (where $M_{\odot}$ is the mass of the sun), and we choose the flow parameters $(\theta, q_{2c}, \gamma)$ as $(35^\circ, 0.20, 4/3)$. Under these parameters, we calculate the total mass outflow rate $R^{\rm{tot}}_{\dot{m}}$ for each point in the parameter space, which is defined as
$$
R^{\rm{tot}}_{\dot{m}}=1-\left[1-R_{\dot{m}}(r_1)\right]\left[1-R_{\dot{m}}(r_2)\right]\left[1-R_{\dot{m}}(r_3)\right].
\eqno(26)
$$
Where, $r_1=0.15$, $r_2=0.25$ and $r_3=0.35$ are three fixed radial distances. We combine the three solutions at these distances to form a point-wise solution for each point in the parameter space, so the total mass outflow rate is composed of the local outflow rates at these positions as shown in equation (26). It is noted that if there is no spiral shock solution for a certain point in parameter space, then the local mass outflow rates have $R_{\dot{m}}(r_1)=R_{\dot{m}}(r_2)=R_{\dot{m}}(r_3)=0$ and thus $R^{\rm{tot}}_{\dot{m}}=0$. It implies that there is no outflow induced by the spiral shock, but it does not exclude other outflows due to different mechanisms.     

In Figure \ref{Figure_7}, we draw the surface projection of the total mass outflow rates $R^{\rm{tot}}_{\dot{m}}$ on the 2-dimensional $P-M_2$ parameter space. In each panel, the color-coded area represents the parameter region with shock-induced outflow, within which $R^{\rm{tot}}_{\dot{m}}$ are indicated with different colors, ranging from $15\%$ to $50\%$, and the blank area is the region without spiral shock and thus without shock-induced outflow. Meanwhile, one can easily observe that the color area expands with the increase of WD mass. This implies that the high-mass WD more easily undergoes the mass outflow than low-mass one. This feature will help to broaden the range of accretion rate for the stable growth of high-mass WDs, which is very narrow for the H-shell burning of WD \citep{Iben-etal89, Wang-18}, through reducing the final accretion rate onto the WD surface. Moreover, it actually overcomes the requirement of the poor assumption of optically thick wind because the excess material is uniformly released during the accretion rather than suddenly released on the WD surface. Another feature stabilizing WD accretion is that the increases of $R^{\rm{tot}}_{\dot{m}}$ with the decrease of $M_2$ exhibited in all of four panels, which balances the increasing mass loss from the companion when its gravitational binding weakens.

After the above analyses on the physical significance of outflow induced by the spiral shock, it is necessary to point out the two limitations in our model. First, due to the self-similarity assumption on the mass density, our model is independent of the practical accretion rate. Indeed, this is not entirely correct for the actual accretion flow, and further study in real-world situations will need to perform the complicated numerical simulation, which is beyond the scope of this paper. However, it would be an interesting future work. Second, for binaries with parameters in the $P$-$M_2$ space shown in Figure \ref{Figure_7}, we can not verify the filling of their donor parts to Roche lobe, but our parameter space includes the regime given by \cite{Wang-etal10} \citep[see also][]{Wang-18} that can eventually evolve to SN Ia. Therefore, the $R^{\rm{tot}}_{\dot{m}}$ in Figure \ref{Figure_7} should be regarded as the possible outflow rate if the accretion disk exists, and it might be applied to some special accretion cases with wind RLOF, which does not need a donor filling the Roche lobe \citep{Mohamed-Podsiadlowski12, Liu-etal17}. 

\section{Discussions and Conclusions}

In this work, we model a non-axisymmetric accretion flow, including spiral shocks around a compact star in a binary system following the previous work of \citet{Chakrabarti-90b} around a single star. Our model incorporates the Roche potential and Coriolis force in the basic equations in a binary system. Moreover, our model can be regarded as a generalized model of \citet{Chakrabarti-90b} (see sub-section 2.5). Here, we calculate the point-wise solutions under the self-similarity assumption, which has been widely used in the studies of accretion theory. We neglect the momentum conservation equation in the vertical direction and consider the flow in vertical equilibrium throughout the disk \citep{Chakrabarti-90b}. The analytical accretion solutions are obtained based on the standard sonic point analysis \citep{Chakrabarti89}. It is observed that the nature of the sonic surfaces and solution topology is determined by the flow parameters \citep{Chakrabarti89, Das-etal01}. We also provide an analytic approach to determine shock locations in the flow.

We find that the flow parameters ($\theta, q_{2c}, \gamma$) and binary parameters ($q, \omega$) play an essential role in determining the nature of the accretion solution. We investigate the spiral shock solutions by supplying the flow and binary parameters and using the analytical expression of shock (equations 22 and 23). It is found that shock-free, as well as shocked solutions, are an essential part of global accretion solution (see Figure \ref{Figure_2}, Figure \ref{Figure_3} and Figure \ref{Figure_4}). We observe that spiral shocks are produced for both the number of shocks $n_s =2$ and 4, depicted in Figure \ref{Figure_2} and Figure \ref{Figure_3}, respectively. The binary parameters have a profound effect on the dynamics of spiral shocks. The strength of shocks is also affected due to the binary parameter in the flow (see Figure \ref{Figure_6}). Further, we compute mass outflow rates from the disk due to the shock compression at the spiral shock using equations (24) and (25). We also investigate the amount of angular momentum dissipated at the shock. Finally, we apply our model to accreting WD in a binary system. We define the total mass outflow rate $R^{\rm{tot}}_{\dot{m}}$ from the local mass outflow rate $R_{\dot{m}}$ (see equation 26) and demonstrates the influence of binary parameters on the $R^{\rm{tot}}_{\dot{m}}$ in the Figure \ref{Figure_7}. According to the applicability of our model, we believe that if the mass outflow induced by spiral shocks is considered in future studies of the progenitor of SN Ia, the requirement of poorly understood assumption of optically thick wind will be overcome. This is particularly beneficial to the mass growth of high-mass WD in SD channel through accreting hydrogen-rich material.

Finally, the present work is based on the second-order approximation of the analytical expression of shock calculations. To get the exact shock locations, we need to incorporate higher-order terms in the shock calculations. So far, we consider an adiabatic accretion flow. In a more realistic accretion flow, we should include viscous dissipation in the flow, and dissipative flow may significantly change the dynamics of the shock \citep{Chakrabarti-Das04, Aktar-etal17}. On the other hand, the magnetized structure of the accretion disk also plays an important role in the WD accretion in a binary system. A complete understanding of the accretion dynamics needs a more realistic investigation by using time-dependent simulation. Moreover, the time-dependent study may be able to explain the time variabilities of the emitted radiation from the disk. This is beyond our scope in the present analysis. Further, we can extend our model to a binary black hole system (XRBs) by considering appropriate gravitational potentials. We will report these aspects elsewhere.

%
\section*{Acknowledgments}
We thank the anonymous referee for very useful comments and suggestions to improve the quality of the paper. This work was supported by the National Natural Science Foundation of China under grants No. 11373002, 11822304, and 12173031.


\bibliography{refs}{} 
\bibliographystyle{aasjournal}

\appendix


\section{Calculation of derivatives at sonic surface}

To obtain $\frac{d q_3}{d \psi}$ at the sonic surface, we apply `l'Hospital rule in equation (13) and, is given by
$$
\left(\frac{d q_3}{d \psi}\right)_c = \frac{\frac{d N}{d \psi}}{\frac{d D}{d \psi}}
\eqno(A1)
$$
Therefore, 
$$
\frac{d D}{d \psi}  =  D_1 \frac{d q_3}{d \psi} + D_2  
\eqno(A2)
$$
where,
\begin{align*}
D_1 &=  \frac{(\gamma +1)}{(\gamma -1)q_3} \left[ - \frac{1}{(\gamma -1)}  - \frac{B^2}{(\gamma -1)}
- \frac{q_w^2 }{2  q_3 }\right] 
\\ D_2 & =  \frac{B}{ q_3}  \frac{(\gamma +1)}{(\gamma -1)}  \left[ \frac{n_\rho + 1}{\gamma} q_3   
+ \frac{q_1^2}{2} + q_2^2  +  \frac{2 \omega  q_2}{\Omega_K} - \alpha_1  \right] + \frac{1}{q_3}  \frac{(\gamma +1)}{(\gamma -1)} \left[- \frac{ q_1 q_2}{2}   -  \frac{2 \omega q_1}{\Omega_K} - \alpha_2    \right] .
\end{align*}
Similarly, we obtain
$$
\frac{d N}{d \psi}  = \left[N_1 - \frac{B}{q_w(\gamma -1)} N_2 - \frac{1}{ q_w(\gamma -1)} N_3\right] \left(\frac{d q_3}{d \psi}\right) + N_2 N_4 + N_3 N_5
\eqno(A3)
$$
where, 
\begin{align*}
N_1 & =  - \frac{(n_\rho +1) B}{\gamma}, ~~~~N_2  = -Bq_1+ \frac{q_2}{2} + \frac{2 \omega}{\Omega_K} + \frac{3}{2} q_w + \frac{3}{2} Bq_1 + \frac{r q_w }{ G} \left(\frac{\partial G}{\partial r} \right) - \alpha_3 B, ~~~ \\ N_3 & = -2Bq_2 - \frac{2 \omega B}{\Omega_K} + \frac{q_1}{2} + \frac{3}{2} q_1 - \alpha_3 + \frac{q_w}{ G} \left(\frac{\partial G}{\partial \phi} \right) , ~~~N_4  =  \frac{1}{q_w} \left[ \frac{n_\rho + 1}{\gamma} q_3  + \frac{q_1^2}{2} + q_2^2  +  \frac{2 \omega  q_2}{\Omega_K} - \alpha_1  \right], ~~~ \\  N_5 &  = \frac{1}{q_w} \left[- \frac{ q_1 q_2}{2}  -  \frac{2 \omega q_1}{\Omega_K} - \alpha_2  \right] .
\end{align*}
Therefore, from Eq. (A1), we get
$$
 \left(\frac{d q_3}{d \psi}\right)_c = \frac{-Q_2 \pm \sqrt{Q_2^2 - 4 Q_1 Q_3}}{2 Q_1}
\eqno(A4)
$$
where,
\begin{align*}
Q_1 &= D_1, ~~~~Q_2 = D_2  - N_1 + \frac{B}{q_w(\gamma -1)} N_2 + \frac{1}{ q_w(\gamma -1)} N_3 ~~~~{\rm ,and}~~~ Q_3 = - N_2 N_4 - N_3 N_5 . 
\end{align*}

\vspace{10 mm}

\section{Calculation of shock location}

$$
\left(\frac{d q_{\parallel}}{d \psi}\right)_c = \frac{1}{q_w \zeta} Q
\eqno(B1)
$$
and 
$$
\left(\frac{d^2 q_{\parallel}}{d \psi^2}\right)_c = \frac{1}{q_w \zeta} \left(\frac{d Q}{d \psi}\right)_c  - \frac{1}{q_w^2 \zeta} \left(\frac{d q_w}{ d \psi}\right)_c Q.
\eqno(B2)
$$
where $q_{\parallel} = \frac{(q_1 -  Bq_2)}{\zeta} $ and $\zeta = (B^2 +1)^{1/2}$.

\begin{align*}
Q& = \frac{(n_\rho +1)}{\gamma} q_3  + \frac{q_1^2}{2}  +  q_2^2  +\frac{2 \omega  q_2}{\Omega_K}  - \alpha_1 + B \frac{q_1q_2}{2} 
+ B \frac{2 \omega q_1}{\Omega_K} + B \alpha_2
\end{align*}
Now, we perform derivative in Eq. (A1) again to get $\frac{d^2 q_3}{d \psi^2}$ at the sonic surface
$$
\left(\frac{d^2 q_3}{d \psi^2}\right)_c = \frac{\left(\frac{d^2 N}{d \psi^2}\right)_c \left(\frac{d D}{d \psi}\right)_c - \left(\frac{d N}{d \psi}\right)_c \left(\frac{d^2 D}{d \psi^2}\right)_c}{\left(\frac{d D}{d \psi}\right)_c^2} = \frac{\left(\frac{d^2 N}{d \psi^2}\right)_c - \left(\frac{d q_3}{d \psi}\right)_c \left(\frac{d^2 D}{d \psi^2}\right)_c}{\left(\frac{d D}{d \psi}\right)_c}
\eqno(B3)
$$
where,
$$
\left(\frac{d^2 N}{d \psi^2}\right)_c =  P_1 \left(\frac{d^2 q_3}{d \psi^2}\right)_c + P_2
\eqno(B4)
$$
and,
$$
\left(\frac{d^2 D}{d \psi^2}\right)_c =  P_3 \left(\frac{d^2 q_3}{d \psi^2}\right)_c + P_4.
\eqno(B5)
$$
Now, substituting Eqs. (B4) and (B5) in Eq. (B3), we get
$$
\left(\frac{d^2 q_3}{d \psi^2}\right)_c = \frac{P_2 -P_4 \left(\frac{d q_3}{d \psi}\right)_c}{\left[\frac{d D}{d \psi} -P_1 + P_3 \left(\frac{d q_3}{d \psi}\right)_c \right]}.
\eqno(B6)
$$
Also, the second derivatives of $q_1$ and $q_2$ are given by
$$
\left(\frac{d^2 q_1}{d \psi^2}\right)_c = P_5 \left(\frac{d^2 q_3}{d \psi^2}\right)_c + P_6
\eqno(B7)
$$
and,
$$
\left(\frac{d^2 q_2}{d \psi^2}\right)_c = P_7 \left(\frac{d^2 q_3}{d \psi^2}\right)_c + P_8
\eqno(B8)
$$
where,
$$
P_9 =-\frac{(\gamma +1)}{(\gamma -1)q_3^2} \left[ - \frac{1}{(\gamma -1)}  - \frac{B^2}{(\gamma -1)}
- \frac{q_w^2 }{2  q_3 }\right]   + \frac{(\gamma +1)q_w^2 }{(\gamma -1)2 q_3^3} ~~~{\rm ,and}~~~
P_{10} = - \frac{(\gamma +1) q_w}{(\gamma -1)q_3^2} \frac{d q_w}{d \psi}
$$

\begin{align*}
P_{11} =- \frac{B (\gamma +1)}{q_3^2 (\gamma -1)} N_4 q_w +  \frac{B (\gamma +1) (n_{\rho} +1)}{q_3 \gamma (\gamma -1)} - \frac{(\gamma +1)}{q_3^2(\gamma -1)} N_5 q_w
\end{align*}
\begin{align*}
P_{12} & =   \frac{B}{ q_3}  \frac{(\gamma +1)}{(\gamma -1)} \left[ q_1  \frac{d q_1}{d \psi} + 2 q_2 \frac{d q_2}{d \psi} + \frac{2 \omega}{\Omega_k} \frac{d q_2}{d \psi}\right] + \frac{1 }{q_3}  \frac{(\gamma +1)}{(\gamma -1)} \left[-\frac{q_2}{2} \frac{d q_1}{d \psi} -\frac{q_1}{2} \frac{d q_2}{d \psi} - \frac{2 \omega}{\Omega_K} \frac{d q_1}{d \psi} \right]
\end{align*}
$$
P_3 = D_1,~~~~P_4 = P_9 \left(\frac{d q_3}{d \psi}\right) ^2 + (P_{10} + P_{11}) \left(\frac{d q_3}{d \psi}\right) + P_{12} ~~{\rm ,and}~~
P_1 =  N_1 - \frac{B}{q_w(\gamma -1)} N_2 - \frac{1}{ q_w(\gamma -1)} N_3
$$
\begin{align*}
P_2 &  = \left[ \frac{d N_1}{d \psi}- \frac{ B}{q_w (\gamma -1)} \frac{d N_2}{d \psi} +\frac{B}{q_w^2 (\gamma -1)} N_2 \frac{d q_w}{d \psi}  
 - \frac{1}{q_w (\gamma -1)} \frac{d N_3}{d \psi} +  \frac{1}{q_w^2 (\gamma -1)} N_3 \frac{d q_w}{d \psi} \right] \left(\frac{d q_3}{d \psi}\right) 
\\ & + N_4 \frac{d N_2}{d \psi} + N_2\frac{d N_4}{d \psi}  + N_5 \frac{d N_3}{d \psi} + N_3 \frac{d N_5}{d \psi} 
\end{align*}
\begin{align*}
P_5 &  =  -\frac{B}{(\gamma-1)q_w}
\end{align*}
\begin{align*}
P_6 &  =  \frac{1}{q_w}\left[ \frac{(n_\rho +1)}{\gamma}  \left(\frac{d q_3}{d \psi}\right)   + q_1  \frac{d q_1}{d \psi}  + 2 q_2 \frac{d q_2}{d \psi}   + \frac{2 \omega}{\Omega_K} \frac{d q_2}{d \psi} \right]  - \frac{1}{q_w^2} \left[\frac{(n_\rho +1)}{\gamma} q_3 - \frac{B}{(\gamma -1)}\frac{dq_3}{d \psi}   + \frac{q_1^2}{2}  + q_2^2  + \frac{2 \omega q_2}{\Omega_K}  - \alpha_1 \right] \frac{d q_w}{d \psi} 
\end{align*}
$$
P_7   =  -\frac{1}{(\gamma-1)q_w}
$$
\begin{align*}
P_8 &  =\frac{1}{q_w} \left[ -\frac{q_2}{2} \frac{d q_1}{d \psi}  -\frac{q_1}{2} \frac{d q_2}{d \psi}  - \frac{2 \omega}{\Omega_K} \frac{d q_1}{d \psi} \right]  -\frac{1}{q_w^2}  \left[ - \frac{q_1 q_2}{2} - \frac{1}{(\gamma -1)} \frac{d q_3}{d \psi} - \frac{2 \omega q_1}{\Omega_K} - \alpha_2 \right] \frac{d q_w}{ d \psi}
\end{align*}

\end{document}